\documentclass[notitlepage,amsmath,superscriptaddress,amssymb,11pt,floatfix,aps,prd,nofootinbib,preprintnumbers]{revtex4-1}
\usepackage{physics}
\usepackage{slashed}
\usepackage{dsfont}
\usepackage{bm}
\usepackage{graphicx}
\usepackage{xcolor}
\usepackage[pdftex,final]{hyperref}
\pdfoutput=1

\newcommand{\be}{\begin{equation}}
\newcommand{\ee}{\end{equation}}
\def\bea{\arraycolsep .1em \begin{eqnarray}}
\def\eea{\end{eqnarray}}
\def\s0#1#2{\mbox{\small{$ \frac{#1}{#2} $}}}
\def\0#1#2{\frac{#1}{#2}}
\newcommand{\muspace}{\mkern1mu}
\renewcommand{\d}{\mathrm{d}}
\newcommand{\psib}{\bar{\psi}}
\newcommand{\la}[1]{\lambda_{#1}}
\DeclareMathOperator{\STr}{STr}
\newcommand{\Nf}{N}

\newcommand{\GammaInt}{\Gamma_k^{\rm int}}
\newcommand{\GammaTwoInt}{\Gamma_{k, \rm int}^{(2)}}

\newcommand{\GammaTwoIntLam}{\Gamma_{\Lambda, \rm int}^{(2)}}
\newcommand{\PS}{S_5}
\newcommand{\ps}{\sigma_5}

\renewcommand{\thesection}{{\bf \Roman{section}}}

\def\step{\\[-1.5ex]}

\begin{document}
\preprint{CERN-TH-2024-078}
\author{Charlie Cresswell-Hogg}
\email{c.cresswell-hogg@sussex.ac.uk}
\affiliation{Department of Physics and Astronomy, University of Sussex, Brighton, BN1 9QH, U.K.}
\author{Daniel F.~Litim}
\email{d.litim@sussex.ac.uk}
\affiliation{Department of Physics and Astronomy, University of Sussex, Brighton, BN1 9QH, U.K.}
\affiliation{Theoretical Physics Department, CERN, 1211 Geneva 23, Switzerland}

\title{Fermions and the Renormalisation Group at Large \emph{N}}

\begin{abstract}
We investigate fermionic quantum field theories using functional renormalisation. In the limit of many fermion flavours $N$, we demonstrate that theories have exact solutions for their quantum effective actions given by quasi-local interaction functionals of fermion bilinears. The structure implies that local potential approximations are exact, exactly solvable, and that field anomalous dimensions vanish. Theories with non-trivial anomalous dimensions may also arise under conditions that are identified. We further demonstrate that higher derivative interactions are inevitably induced by point-like ones, including at large-$N$. The local potential flows for fermionic theories with the most general  $U(N)$ symmetric interactions are provided. For sample theories with scalar, pseudo-scalar, vector, or axial-vector interactions, we identify conformal fixed points, scaling dimensions, conformal manifolds, and quantum-induced shifts in  scaling dimensions of higher derivative interactions. We also study fermion mass generation, and subleading modifications due to finite $N$ corrections. Implications for conformal field theories, and applications in condensed matter and particle physics are indicated.
\end{abstract}

\maketitle

\tableofcontents

\newpage

\section{\bf Introduction}

Strongly interacting  fermions play an important role in condensed matter and particle physics, covering  diverse phenomena such as 
symmetry breaking, dynamical generation of  mass,  formation of bound states,    low-energy descriptions of the strong nuclear force, and critical points with quantum, topological, or thermal phase transitions~\cite{Metzner:2011cw,Braun:2011pp}. 
Relativistic fermions   also appear prominently  
in models characterising  Dirac materials  \cite{Rosenstein:1993zf,Herbut:2006cs,Hands:2008id,Herbut:2009qb,Vafek:2013mpa,Wehling:2014cla,Classen:2015ssa,Boyack:2020xpe,Parthenios:2023apm,Biedermann:2024zop}. 
Critical fermions  may  become 
 non-perturbatively renormalisable \cite{Wilson:1972cf,Parisi:1975im,Eguchi:1976iz,Shizuya:1979bv,Gawedzki:1985ed,Rosenstein:1988pt,Kikukawa:1989fw,deCalan:1991km,Hands:1991py,Hands:1992be,Gracey:1993kc,Gies:2010st,Jakovac:2014lqa,Gehring:2015vja,Dabelow:2019sty,Cresswell-Hogg:2022lgg,Cresswell-Hogg:2022lez} in the spirit of  asymptotic safety  \cite{Weinberg:1980gg,Reuter:1996cp,Reuter:2001ag,Litim:2003vp,Gies:2010arw,Braun:2010tt,Litim:2014uca,Bond:2016dvk,Bond:2018oco,Gies:2024ugz}, 
and   are  of interest for the  conformal bootstrap  \cite{Poland:2018epd} or as gravitational duals under  the AdS/CFT conjecture, e.g.~\cite{Giombi:2012ms}.
 \step

A powerful   continuum method in the study of strongly coupled phenomena is provided by   functional renormalisation  \cite{Polchinski:1983gv,Wetterich:1992yh,Ellwanger:1993mw,Morris:1993qb}. 
It is based on a wilsonian momentum cutoff \cite{Wilson:1971bg,Wilson:1971dh} to facilitate the  integrating-out of momentum modes,  and interpolates between  a microscopic action at short distances and  the full quantum effective action once all fluctuations are integrated out. 
The method has  been applied  to a wide range of non-perturbative phenomena \cite{Dupuis:2020fhh} including  fermionic systems, e.g.~\cite{Comellas:1995ea,Jungnickel:1995fp,Gies:2001nw,
Gies:2002hq,Gies:2003dp,
Pawlowski:2005xe,
Metzner:2011cw,Braun:2011pp,
Aoki:2013gda,Jakovac:2013jua,Aoki:2014ola,Jakovac:2014lqa,
Vacca:2015nta,Classen:2015ssa,Parthenios:2023apm,Gies:2023cnd},  often with the help of 
auxiliary fields from  dynamical bosonisation  \cite{Gies:2001nw,Pawlowski:2005xe} or Hubbard-Stratonovich transformations \cite{Hubbard:1959ub,Stratonovich}.\step

The  aim of this paper is to combine functional renormalisation for fermions with the large-$N$ limit, with $N$ being the number of particle species or flavours. 
The reason for doing so is twofold. Firstly, large-$N$ limits  often provide rigorous  control over  fluctuations and critical points including beyond  perturbation theory, and enable first principle insights  that otherwise are difficult to achieve \cite{Moshe:2003xn}. 
  Secondly,   it was noticed early on \cite{Wegner:1972ih,Ma:1973zu,Nicoll:1976ft,CHANG1992279} that a large-$N$ limit in combination with functional renormalisation renders  local potential approximations of scalar  theories  exact. Most notably,  
  the large-$N$ functional form of quantum effective actions   has also been identified  \cite{DAttanasio:1997yph}. 
  Given the potential relevance of these  insights beyond scalar theories, 
  and as a matter of principle, it is crucial to demonstrate  explicitly that similar  simplifications    arise in  fermionic theories.\step
 
 In this spirit, we put forward a  large-$N$ study of general fermionic theories under the prism of functional renormalisation. 
 To achieve our  results,  we work directly in terms of the elementary fermionic degrees of freedom, rather than 
  introducing collective fields early on, and in terms of suitable fermion bilinears $J$  
   reflecting the underlying Clifford algebra structure. We will then demonstrate that the exact functional flow for a non-perturbative quantum effective action $\Gamma_k [ \psi, \psib ]$ has exact functional solutions of the form
  \be\label{eq:GammaFormF}
\Gamma_k [ \psi, \psib ] = \int_x \psib_i ( x ) \slashed{\partial} \mkern1mu \psi_i ( x ) + F_k [J] \,,
\ee
up to  subleading corrections of order $1/N$. Here, $k$ denotes the  RG momentum scale ($k\to 0$ in the physical limit) 
and $F_k$  a local functional depending on the set of  flavour-singlet bilinears~$J$  and derivatives thereof. 
The  result \eqref{eq:GammaFormF} offers a substantial simplification over the structure of effective actions at finite $N$, with the added benefit that corrections are $1/N$ suppressed. We also demonstrate that  the  flow for the local potential interactions -- the part of the functional $F_k [ J ]$ that is independent of derivatives --  is large-$N$ exact  for any and all subsets of bilinears, that it can be solved exactly,  that Fierz ambiguities are absent, and  identify conditions under which fermionic theories  do {\it not} take the form \eqref{eq:GammaFormF} including at large-$N$.\step

We  then put our method to work and provide exact functional flows  in the local potential approximation for  theories in general dimensions, and with the most general  interactions with a global $U(N)$ symmetry.  Moreover, we apply our method to a range of sample fermionic theories with scalar, pseudo-scalar, vector, or axial-vector interactions, to  determine  interacting fixed points and global fixed point solutions, universal scaling dimensions, and conformal manifolds, if available. We also demonstrate that higher derivative interactions are invariably induced by point-like interactions, and exemplarily compute the quantum-induced shifts to their scaling dimensions. Finally, we revisit aspects of  fermion mass generation to show  that this phenomenon is at least $1/N$ suppressed if mass is not protected by a symmetry,  irrespective of interactions.\footnote{This generalises a curious  feature observed   in 3d Gross-Neveu theories with six-fermion interactions  \cite{Cresswell-Hogg:2022lgg,Cresswell-Hogg:2022lez}.}\step

The paper is organised as follows. 
Sec.~\ref{sec:RG} introduces the basics of functional renormalisation  for fermions  (Sec.~\ref{sec:fermionicRG}),  and provides exact  flows  in the limit of many fermion flavours $N$ (Sec.~\ref{sec:exactSols}). Conditions under which   large-$N$  flows  for pointlike interactions become exact, and exactly solvable,  are detailed (Sec.~\ref{sec:LPA}), and general local potential flows for theories with global $U(\Nf)$ symmetry are provided (Sec.~\ref{sec:localPotentials}). 
Sec.~\ref{sec:applications} covers applications of our methods to find conformal fixed points and scaling dimensions in various fermionic theories including scalar and pseudo-scalar interactions (Sec.~\ref{sec:SandP}), vector and axial-vector interactions (Sec.~\ref{sec:VandA}), and higher derivative interactions (Sec.~\ref{sec:higherDerivatives}),  also covering aspects of fermion mass generation (Sec.~\ref{sec:fermionMass}). 
We conclude in Sec.~\ref{sec:conclusion}.  
Appendices deal with technical aspects such as the completeness of the pointlike interactions basis (App~\ref{app:2nFinteractions}), 
and Dirac algebra conventions in euclidean signature (App.~\ref{sec:Dirac}).\step

\section{\bf Renormalisation Group}
\label{sec:RG}

In this paper, we are concerned with quantum field theories of $\Nf$ self-interacting Dirac fermions, transforming in the fundamental representation of a global $U(\Nf)$ flavour symmetry. We work in $d$ euclidean dimensions and denote the number of spinor components per fermion flavour by $d_\gamma$. The dynamics of such theories are conveniently encoded in the quantum effective action $\Gamma$, the generating functional of 1PI correlation functions. Perturbatively, the effective action can be computed from 1PI diagrams, however the essential physics of fermionic systems is oftentimes strongly coupled and methods besides weak-coupling perturbation theory are needed to capture it. One such method is the functional RG \cite{Wetterich:1992yh,Ellwanger:1993mw,Morris:1993qb}. It provides a framework to compute the quantum effective action as the solution of an exact functional differential equation, derived directly from the path integral, and without reference to a perturbative expansion. 

\subsection{Functional Renormalisation for Fermions}
\label{sec:fermionicRG}

The technique of functional renormalisation is based on the introduction of 
a momentum cutoff  into the path integral definition of quantum or statistical field theory \cite{Wilson:1971bg,Polchinski:1983gv,Wetterich:1992yh}. For a general fermionic theory in euclidean space-time, we introduce the partition function as
\be
\exp W_k [ \eta, \bar\eta ] = \int {\cal D} \psi \mkern2mu {\cal D} \psib \, \exp{ -S [ \psi, \psib ] - \psib \cdot R_k \cdot \psi + \bar\eta \cdot \psi + \psib \cdot \eta} \,.
\ee
 Here $S [ \psi, \psib ] $   denotes the  fundamental action, $R_k$ the wilsonian regulator function,  $k$  the corresponding RG momentum scale, $W_k$  the coarse-grained Schwinger functional, and dots stand for summation over discrete indices as well as integration over common arguments in position or momentum space. 
The Schwinger functional $W_k$ relates to the  quantum effective action $\Gamma_k$  by a  Legendre transform. 
The regulator $R_k$ is a two-point function which enacts a scale-dependent coarse-graining by suppressing the propagation of  modes with momenta below the scale $k$. 
As a function of momenta $q^2$, it  obeys $R_k(q) > 0$ for $q^2/k^2 \to 0$ to  suppress the propagation of low-momentum modes, and $R_k(q)\to 0$ for $k^2/q^2 \to 0$ to remove the cutoff in the IR.\step

Most importantly, the cutoff term 
induces a  scale-dependence $\partial_t\equiv k\partial_k$, which takes  the form of  an exact functional flow for  the quantum effective  action
\cite{Wetterich:1992yh,Ellwanger:1993mw,Morris:1993qb}
\be\label{eq:functionalFlowMain}
\partial_t \Gamma_k = \frac12 \STr \left\{ \big[ \Gamma^{(2)}_k + R_k \big]^{-1} \cdot \partial_t R_k \right\} \, .
\ee
Here, $\Gamma^{(2)}_k$ denotes the functional Hessian of the effective action with respect to all fields, while the supertrace operation $\STr$ stands for a functional trace 
over all continuous and discrete indices, including relative signs as appropriate for anticommuting fields.
Subject to suitable initial conditions, 
 $\Gamma_k$ interpolates between the microscopic theory at the high scale ($k\to \Lambda$) and the full quantum effective action $(k\to 0)$.  
   Further, the   flow \eqref{eq:functionalFlowMain} is IR finite by virtue of the regulator term in the denominator, and UV finite owing to the suppression generated by
$\partial_t R_k ( q )$ \cite{Litim:1998nf}. \step

The functional flow  \eqref{eq:functionalFlowMain} is closely linked to other exact functional differential equations such as the Callan-Symanzik equation  in the limit where the regulator is reduced to a mass term, and the Polchinski equation \cite{Polchinski:1983gv} by means of a functional Legendre transformation \cite{Litim:2005us,Morris:2005ck,Litim:2018pxe}. 
At weak coupling, iterative solutions of the flow generate  perturbation theory to all  loop orders  \cite{Litim:2001ky,Litim:2002xm}.  At strong coupling,  non-perturbative  approximations such as the derivative expansion, vertex expansions, or mixtures thereof are available  \cite{Litim:1998nf,Freire:2000bq,Pawlowski:2005xe,Dupuis:2020fhh}. The stability and convergence of approximations such as the derivative expansion can be controlled as well, e.g.~\cite{Litim:2001fd,Litim:2001dt,Litim:2010tt,Balog:2019rrg,Baldazzi:2023pep}. Functional flows for fermions with and without  the help of collective fields and bosonisation have been developed within the Polchinski \cite{Comellas:1995ea} and Wetterich versions of the flow, e.g.~\cite{Bornholdt:1992za,Jungnickel:1995fp,Gies:2001nw,Jaeckel:2002rm,Gies:2002hq,Gies:2003dp,Pawlowski:2005xe,Braun:2011pp,Metzner:2011cw,Aoki:2013gda,Jakovac:2013jua,Aoki:2014ola,Jakovac:2014lqa,Vacca:2015nta}. 
In this work, we study  functional flows  for fermions directly in terms of the elementary fermion fields, in analogy to the treatment of scalar fields, and without the  use of composite fields. \step

One of the  technical challenges when dealing with fermions instead of  scalars relates to the additional matrix structure at the level of the functional Hessian \eqref{eq:functionalFlowMain}, 
where all four second-order derivatives with respect to the fields $\psi$ and $\psib$ must be included 
to account for interactions. 
 In order to simplify this structure, we  work  in a ``doubled representation'' and combine $\psi$ and $\psib$ into a single field,
\be\label{eq:chiDef}
\chi =
\begin{pmatrix}
\psi \\
\psib^T
\end{pmatrix} \,.
\ee
Notice that the Grassmann-odd components $\chi_a$ treat  spinor and flavour indices  as a single field-space index $a$ running from 1 to $2 d_\gamma \Nf$. 
The new basis is  particularly well-suited for a large $N$ study of fermionic theories, because it enables a direct generalisation of  arguments originally  made for large $N$ scalar  theories~\cite{DAttanasio:1997yph}.
In the doubled basis \eqref{eq:chiDef},  kinetic terms are written  as 
\be\label{kinetic}
\int_x \psib_i ( x ) \,\gamma^\mu \partial_\mu \,\psi_i ( x ) = \frac12 \int_x \chi_a ( x )\, \mkern1mu \Gamma_{ab}^\mu \mkern1mu \partial_\mu \,\mkern1mu \chi_b ( x )
\ee
with the help of  matrices $\Gamma^\mu$, which are  given by
\be
\Gamma^\mu = 
\begin{pmatrix}
0 & (\gamma^\mu)^T \otimes \mathds{1}_{\Nf}
\\
\gamma^\mu \otimes \mathds{1}_{\Nf} & 0
\end{pmatrix}\,.
\ee
 The matrices  $\Gamma^\mu$ are symmetric in their field-space indices $a$ and $b$ 
owing to the fact that integration by parts is used to achieve \eqref{kinetic}. It follows that the contraction $\chi_a\, \Gamma^\mu_{ab}\, \chi_b \equiv 0$  vanishes identically  for any Grassmann-odd $\chi_a$ at the same spacetime point. This property automatically implements some of the nontrivial cancellations which occur 
when working in terms of the $(\psi,\psib)$ blocks individually \cite{Jakovac:2013jua}.\step

Besides kinetic terms, quantum effective actions  $\Gamma_k $ in general  also depend on fermion bilinears such as
\be\label{eq:JAdefPsi}
J^A ( x ) \equiv \psib_i ( x ) \mkern2mu \gamma^{(A)} \mkern1mu \psi_i ( x ) \, ,
\ee
where $i$ is the flavour index in the fundamental representation and $\gamma^{(A)}$ are a set of Dirac matrices spanning the relevant Clifford algebra. In fact, {\it any} pointlike interaction can be written exclusively in terms of a set of flavour-singlet bilinears \eqref{eq:JAdefPsi}, as demonstrated in App.~\ref{app:2nFinteractions}.\step

 In the case where the basic fermions are  four-component Dirac spinors, as adopted here,  a complete basis in four dimensions  ($\mu = 0, 1, 2, 3$) contains 16 independent  invariants, which we take to be 
\be\label{eq:DiracBasis4d}
\gamma_{\rm 4d}^{(A)} \in \left\{ \mathds{1}, \gamma^\mu, \gamma^5,  i \gamma^\mu \gamma^5, i \gamma^{\mu\nu}\ 
\right|\left. {\rm for}\ {0\le\mu < \nu\le 3} \right\} \, ,
\ee
with $\gamma^{\mu\nu} = \frac12 [ \gamma^\mu, \gamma^\nu ]$. Incidentally, the same type of basis can also be used in three dimensions. There, the four-component spinor representation is reducible, the spacetime index $\mu$ runs from 0 to 2, and the leftover matrix $\gamma^3$ in \eqref{eq:DiracBasis4d} should  be treated in analogy to $\gamma^5$, once more giving 16 independent  invariants,
\be\label{eq:DiracBasis3d}
\gamma_{\rm 3d}^{(A)} \in \left\{ \mathds{1}, \gamma^\mu,\gamma^3, \gamma^5,  i \gamma^\mu \gamma^3,i \gamma^\mu \gamma^5, i\gamma^3 \gamma^5, i \gamma^{\mu\nu}
\ \right|\left. {\rm for}\ {0\le\mu < \nu\le 2} \right\} \,.
\ee
Note that Lorentz covariance is not manifest on the level of bilinears \eqref{eq:JAdefPsi}. Rather, it restricts the allowed combinations in which the bilinears $J^A$ may arise in the effective action.\step

 In order to express fermion bilinears and their interactions 
 in the doubled basis \eqref{eq:chiDef}, it is convenient to  introduce the matrices
\be\label{eq:M}
M[X] \equiv 
\begin{pmatrix}
0 & -X^T \otimes \mathds{1}_{\Nf}
\\
X \otimes \mathds{1}_{\Nf} & 0
\end{pmatrix} \,.
\ee
In terms of these, we can write a general fermion bilinear \eqref{eq:JAdefPsi} in the form
\be\label{def-JA}
J^A ( x ) = \frac12 \, \chi_a ( x ) \, M^A_{ab} \ \chi_b ( x ) \,,
\ee
where we recall that the sum over repeated indices $a, b, \cdots$ runs from $1$ to $2 d_\gamma \Nf$. Notice that the symplectic matrices $M^A\equiv M [ \gamma^{(A)} ]$ are anti-symmetric in their field-space indices $a$ and $b$. 
They  decompose naturally  into blocks corresponding to the $\psi$ and $\psib$ subspaces, with each block having size $d_\gamma \Nf \times d_\gamma \Nf$, confirming that $M^A \in Sp ( 2 d_\gamma \Nf , \mathbb{C} )$. 
It is also convenient to introduce the matrix $\Omega= M [ \mathds{1} ]$, which  represents a symplectic form on  field space and satisfies $\Omega^2_{ab} = -\delta_{ab}$. 
Finally, we note that the matrices $M^A$ and $\Gamma^\mu$ are all traceless matrices and that the product of $\Omega$ and $\Gamma^\mu$ with respect to either left or right multiplication inherits a Clifford algebra structure from $\gamma^\mu$,
\be\label{eq:GammaClifford}
\left\{ \Omega \,\Gamma^\mu, \Omega\, \Gamma^\nu \right\} = \left\{ \Gamma^\mu\, \Omega, \Gamma^\nu \,\Omega \right\} = 2 \mkern1mu \delta^{\mu\nu} \mkern1mu\, \mathds{1}_{2 d_\gamma \Nf} \,.
\ee
With these considerations in place, our setup allows for the  study of general fermionic quantum field theories including at large $N$, to which we turn next.

\subsection{Exact Functional Solutions  at Large \emph{N}}
\label{sec:exactSols}

Next, we demonstrate 
that the  functional flow  \eqref{eq:functionalFlowMain} admits exact  solutions of the form \eqref{eq:GammaFormF} at large $N$. 
To achieve the result, we take advantage of the choice of basis \eqref{eq:chiDef}, \eqref{kinetic} together with a finite set of independent flavour-singlet fermion bilinears  \eqref{eq:JAdefPsi}, \eqref{def-JA}, and the one-loop structure of the  exact functional flow \eqref{eq:functionalFlowMain}. 
We also  benefit from structural  insights achieved 
in the context of scalar theories whose  interactions  are functionals of a single scalar field current $J^\phi=\phi_i\phi_i$    \cite{DAttanasio:1997yph}. \step

To these ends, it is convenient to split the effective action into a free part and an interaction functional $\GammaInt$ as 
\be\label{eq:GammaBarSplit}
\Gamma_k [ \chi ] = \frac12 \int_x \chi_a ( x ) \muspace \Gamma^\mu_{ab} \muspace \partial_\mu \chi_b ( x ) + \GammaInt [ \chi ] \, ,
\ee
which, thus far, is still completely general. For the purposes of our analysis, 
mass terms and corrections to the kinetic term 
are treated as interactions. Since the free part is RG-scale independent, we recast the flow equation for $\Gamma_k$ as a flow for the interactions,
\be\label{eq:flowGammaBar}
\partial_t \GammaInt = \frac12 \STr \left\{ \big[ 1 + \Delta_k \cdot \GammaTwoInt \big]^{-1} \cdot \Delta_k \cdot \partial_t \Delta_k^{-1} \right\} ,
\ee
where $\Delta_k$ stands for the massless free propagator in the presence of the regulator $R_k$. As a differential operator,
\be\label{eq:Delta}
\Delta_k^{-1} = \Gamma^\mu \partial_\mu + R_k ( -i \partial )\,.
\ee 
We employ a notation where two-point functions are treated as matrices in position or momentum space and dots signify ``matrix multiplication'' with respect to both discrete and continuous indices. The quantity $\GammaTwoInt$ denotes the Hessian with respect to $\chi$ of the interaction part of the action,
\be\label{eq:HessianDef}
\big[ \GammaTwoInt \big]_{ab} ( x, y ) = \frac{\roarrow \delta\GammaInt \loarrow\delta}{\delta \chi_a ( x ) \, \delta \chi_b ( y )} \, .
\ee
Let us reiterate that these considerations are fully general and  applicable for any  $N$. \step

Simplifications occur at large $N$  which render the flow  exactly solvable, as we now discuss.
The proof of our claim proceeds by first {\it assuming} that solutions take the form \eqref{eq:GammaFormF} at a reference scale $k = \Lambda$, and then establishing that this form is preserved by the flow at \emph{all} scales, in the limit $\Nf \to \infty$. Of course, the ``assumption'' can always be satisfied by taking the initial condition for the flow to be of the required form. To that end, we take the interaction part of the effective action at $k = \Lambda$ to be a local functional of the bilinears $J^A$, by which we mean it can be written as the integral of a function of $J^A ( x )$ and their derivatives at a single spacetime point,\footnote{We allow derivatives of arbitrarily high order. Functionals of this kind are referred to as ``perturbatively local''~\cite{Weinberg:1997rv}.}
\be\label{eq:Initial}
\Gamma^{\rm int}_{k = \Lambda} [ \chi ] = F_\Lambda [ J ] \, ,
\ee
where $J$ collectively refers to the set of  bilinears \eqref{eq:JAdefPsi}, \eqref{eq:DiracBasis4d}, and the scale $\Lambda$ can be viewed as  the high  scale. Then the Hessian \eqref{eq:HessianDef} decomposes into two terms,
\be\label{eq:d2GammaBar}
\big[ \GammaTwoIntLam \big]_{ab} ( x, y ) = \sum_A \delta ( x - y ) \muspace M^A_{ab} \mkern2mu \frac{\delta F_\Lambda}{\delta J^A ( x )} - \sum_{A,B} \muspace \xi^A_a ( x ) \muspace \xi^B_b ( y ) \mkern2mu \frac{\delta^2 F_\Lambda}{\delta J^A ( x ) \delta J^B ( y )} \, ,
\ee
where we introduced the variables $\xi^A_a ( x ) = M^A_{ab} \mkern1mu \chi_b ( x )$, also using \eqref{eq:M}. 
Notice that we perform a conventional large $\Nf$ limit whereby the effective action scales with $\Nf$, the fermion fields scale with $\sqrt{\Nf}$, and the Hessians 
remain of order unity, independent of $\Nf$. The form of the Hessian  \eqref{eq:d2GammaBar} implies that contributions to the RG flow $\propto \Nf$ originate from the term $\propto M\delta F/\delta J$  upon inversion and tracing in \eqref{eq:flowGammaBar}. In comparison, the second term $\propto\delta^2 F/\delta J\delta J$  only provides contributions to the flow that are of order unity, and $1/N$ subleading   in comparison to the first term. \step

This result can be understood by expanding the functional inverse on the right hand side of \eqref{eq:flowGammaBar} in a formal power series,
\be\label{eq:GammaBarSeries}
\left( 1 + \Delta_\Lambda \cdot \GammaTwoIntLam \right)^{-1} = \sum_{n=0}^\infty \left( -\Delta_\Lambda \cdot \GammaTwoIntLam \right)^n \, ,
\ee
where we recall that $\GammaTwoIntLam = {\cal O} ( N^0 )$, meaning that all terms in the series contribute at the same order in $N$. 
Next, note that all terms in the series involving the second-derivative part of the decomposition \eqref{eq:d2GammaBar} are of the form 
$X \cdot \xi \muspace \xi^T \cdot Y$, 
for some Grassmann-even two-point functions $X$ and $Y$. This remains the case when including the regulator terms in \eqref{eq:flowGammaBar}. Then, taking the trace yields
\be\label{eq:subleadingTr}
\Tr ( X \cdot \xi \muspace \xi^T \cdot Y ) = - \xi^T \cdot Y \cdot X \cdot \xi,
\ee
without any factors of $\Nf$. This should be contrasted with contributions stemming  from the first-derivative term in \eqref{eq:d2GammaBar}, which can yield explicit factors of $\Nf$ because products of the field space matrices $M^A$ and $\Gamma^\mu$ may take the form $X \otimes \mathds{1}_{2 \Nf}$, where $X$ is a product of Dirac matrices.\step

From the above considerations, it follows that all large-$N$ leading contributions to the flow \eqref{eq:flowGammaBar} are contained in
\be\label{eq:largeNflowGammaBar}
\partial_t \GammaInt \equiv \partial_t F_k = \frac12 \STr \mkern-2mu \left( Q^{-1} \cdot \Delta_k \cdot \partial_t \Delta_k^{-1} \right) \, ,
\ee
where
\be\label{eq:Q}
Q_{ab} ( x, y ) = \delta_{ab} \mkern1mu \delta ( x - y ) + [ \Delta_k ( x - y ) ]_{ac} \mkern2mu M^A_{cb} \mkern2mu \frac{\delta F_k}{\delta J^A ( y )} \, .
\ee
The key observation in the proof of our claim is now that the right hand side of this equation is a functional only of the bilinears $J^A$. Therefore, integrating \eqref{eq:largeNflowGammaBar} in RG time, we see that the form 
$\GammaInt [ \chi ] = F_k [ J ]$ 
is preserved at every infinitesimal RG step, and by extension for all $k$, provided that the initial condition for the flow (at the arbitrary scale $k = \Lambda$) is of the same form. Therefore, at leading order in large $N$, a class of exact solutions of the functional flow \eqref{eq:functionalFlowMain} is given by
\be\label{eq:GammaExactSol}
\Gamma_k [ \chi ] = \frac12 \int_x \chi_a ( x ) \muspace \Gamma^\mu_{ab} \muspace \partial_\mu \chi_b ( x ) + F_k [ J ] \, ,
\ee
for all RG schemes and with $F_k$ satisfying \eqref{eq:largeNflowGammaBar}.  
The result \eqref{eq:GammaExactSol} is central for what follows and derserves a few remarks.
\begin{itemize}

\item[(i)] {\it Anomalous dimensions} \\
The exact  quantum effective action $\Gamma_k$, at large $N$,  consists of two terms, the classical fermion kinetic term, and a general functional $F_k[J]$ of the fermion bilinears $J$. Therefore, quantum corrections to the fermion kinetic term, if they exist, must be contained in the second term. However, 
 the only terms in $F_k[J]$ which are both bilinear in the fields and contain derivatives are  total derivatives, which do not contribute to the physics.  Consequently, for theories with \eqref{eq:GammaExactSol} there is no wave function renormalisation at large $N$ and the fermion anomalous dimension vanishes identically, $\eta_\psi \equiv 0$. 
 
\item[(ii)] {\it Local potential approximation} \\
The second term in \eqref{eq:GammaExactSol}, the functional $ F_k [ J ]$,  contains derivative and non-derivative interactions. The non-derivative interactions can be written as $F^{\rm LPA}_k [ J ]= \int_x V_k ( J ( x ))$ in terms of  a local potential function $ V_k ( J ( x ))$. The local potential approximation (LPA for short) consists in replacing $F_k[J]$ in \eqref{eq:GammaExactSol} by $F^{\rm LPA}_k [ J ]$. Most importantly, the local potential approximation becomes exact at large $N$, meaning that the flow of the local potential $\partial_t V_k$, as extracted from \eqref{eq:largeNflowGammaBar}, solely depends on $V_k$ without receiving corrections from higher derivative interactions 
(see Sec.~\ref{sec:LPA}). Consequently, $V_k$ can be determined exactly. 

\item[(iii)]  {\it Derivative interactions} \\
The second term in \eqref{eq:GammaExactSol} also contains  derivative interactions, $F^{\rm deriv}_k [ J ]= F_k[J] -F^{\rm LPA}_k [ J]$. However, the result  \eqref{eq:GammaExactSol} dictates that  derivative interactions are restricted to those that can be built out of  fermion bilinears $J$ and derivatives thereof, whereas any other types of derivative interactions 
are suppressed parametrically, and at least as $1/N$. As such, the large-$N$ limit entails an infinite reduction in the number of derivative operators in the quantum effective action. We further emphasise that non-derivative interactions in $F^{\rm LPA}_k [ J ]$  act as sources for derivative interactions $F^{\rm deriv}_k [ J ]$. We conclude that derivative interactions are in general    switched-on by fluctuations including at large $N$ (for an explicit example, see Sec.~\ref{sec:higherDerivatives}). 

\item[(iv)] {\it Fierz ambiguities} \\ 
The matrices $\gamma^{(A)}$ in \eqref{eq:JAdefPsi} 
span the Clifford algebra to ensure that all possible pointlike interactions with global $U(\Nf)$ symmetry are included in the analysis. 
This choice, however, does neither   enter the derivation of the large $N$ flow \eqref{eq:largeNflowGammaBar}, \eqref{eq:Q} nor the result \eqref{eq:GammaExactSol}. 
As a direct corollary, 
and provided that $F_\Lambda[J]$ at the high scale $\Lambda$, \eqref{eq:Initial}, only depends 
 on a subset $\{J\}$ of the fermion bilinears \eqref{eq:JAdefPsi}, 
 it follows that the RG flow will not generate dependences on any of the  omitted bilinears. 
 In other words, in the large $N$ limit the RG flows are closed for any subset  $\{J\}$.\footnote{This is consistent with, and generalises, observations from functional RG studies truncated at the four-fermion level \cite{Gehring:2015vja} as well as from auxiliary-field methods \cite{Weinberg:1997rv}.} We conclude that Fierz ambiguities, which may arise if an incomplete set of interactions is considered  \cite{Jaeckel:2002rm,Jaeckel:2003uz}, are suppressed in the large $N$ limit.

\item[(v)] {\it Subleading corrections and $\eta_\psi\neq 0$} \\
At finite $N$, contributions from  the subleading terms $\propto \delta^2F_k/\delta J^A\delta J^B$ in \eqref{eq:d2GammaBar} can no longer be neglected. 
These additional terms involve  tensor products $\xi_a^A \xi_b^B$ which under a trace such as in \eqref{eq:subleadingTr} generate new  types of derivative interactions in \eqref{eq:GammaExactSol}, different from those contained in $F_k[J]$. 
 In particular, new derivative interactions involving  $\psib \slashed{\partial} \psi$ are generated when tracing over combinations of tensor products of the fields and the propagator $\Delta_k$. It follows that theories genuinely develop 
 non-trivial fermion anomalous dimensions, $\eta_\psi\neq 0$, and  that the local potential approximation ceases to be exact. For the same reasons, the large-$N$ closure of Fierz-incomplete flows does not survive at finite $N$, and  complete sets of symmetry-compatible bilinears must be retained instead. We conclude that  none of the features (i) - (iv)  persist at finite $N$, even though modifications are $1/N$ subleading. 
As a final remark, we stress that  fermions may  achieve non-vanishing anomalous dimensions including at infinite $N$, and despite of (i). The necessary and sufficient condition for this to occur is that interactions are not of the form \eqref{eq:Initial}. 
\end{itemize}

With these results at hand, and together with the findings  of  \cite{DAttanasio:1997yph}, it is now evident that and how  \eqref{eq:GammaExactSol}  generalises to theories with both scalar and fermionic degrees of freedom. 
Whenever the number of fermionic degrees of freedom is parametrically larger than the number of scalar degrees of freedom, 
quantum effective actions will display, in addition to classical kinetic terms for all fields, a functional $F_k$ that  depends on all possible scalar invariants and fermionic singlet bilinears for which all of the above points (i) - (v) are applicable, and valid for each and any subset of bilinears. 
Anomalous dimensions of scalar fields may be nonvanishing in this case. 
Depending on the specifics of global symmetries, the results also generalise similarly to the case of many fermions coupled to many scalar fields.

\subsection{Local Potential Approximation}
\label{sec:LPA}

 We now turn to fermionic theories with microscopic interaction functionals \eqref{eq:Initial}  in the local potential approximation. 
The latter consists in approximating the quantum effective action $\Gamma_k$ by the sum of a classical kinetic term and a scale-dependent effective potential $V_k$ for bilinears in the fields \cite{Wegner:1972ih,Nicoll:1974zz,Golner:1985fg,Ball:1994ji,Litim:1995ex,Tetradis:1995br,Comellas:1997tf,DAttanasio:1997yph,Morris:1997xj}. As such, the LPA encompasses mass terms as well as all  one-particle irreducible $n$-point functions at vanishing momenta. It  is termed local in that $V_k$  constitutes an ordinary function of the fields at a single spacetime point, without derivatives. The aim of  this section is to demonstrate  that the RG flow of local potential interactions for $U(N)$ symmetric fermionic theories becomes exact  in the large $N$ limit,  and exactly solvable. Our proof benefits from the field basis introduced in \eqref{eq:chiDef} with \eqref{kinetic}, \eqref{def-JA}, and also utilises observations made previously 
in the context of scalar  theories  \cite{DAttanasio:1997yph}.\step

For the fermionic theories discussed here, the local potential approximation takes the form
\be\label{eq:LPAinteractions}
\GammaInt [ \chi ] \approx \int_x V_k ( J ( x ) ) \equiv F^{\rm LPA}_k[J]\,,
\ee
and assumes that the quantum effective action is well-approximated by  local potential interactions, and that derivative interactions are of subleading relevance.  However, even at large $N$, 
the interactions included in \eqref{eq:LPAinteractions}  cover only a subset of those  in the exact solution $\GammaInt [ \chi ] =F_k[J]$, which consists of a local effective potential part $F^{\rm LPA}_k[J]$ and of a part involving derivative interactions, $F^{\rm deriv}_k[J]=F_k[J]-F^{\rm LPA}_k[J]\neq 0$. Clearly, a local potential  approximation  \eqref{eq:LPAinteractions} never represents the entirety of the quantum effective action.\footnote{See Sec.~\ref{sec:higherDerivatives} for an explicit computation  of interactions in $F^{\rm deriv}_k[J]$ and why they are genuinely generated  by pointlike interactions.} \step

With this disclaimer in mind, we now demonstrate that the approximation \eqref{eq:LPAinteractions}  nevertheless becomes {\it exact} in the large $N$ limit, provided interactions are general functionals of fermionic bilinears. What is meant by this statement is that the effective potential $F^{\rm LPA}_k[J]$ can be determined exactly, without any approximations, and the reason why this has become a possibility  at large $N$ is that higher derivative interactions contained in  $F^{\rm deriv}_k[J]$ no longer couple   back into the  flow of the local potential $\partial_t F^{\rm LPA}_k[J]$. In other words, the non-perturbative functional RG flow for the local potential  is closed and only driven by the local potential interactions themselves, and which therefore can  be integrated exactly without the necessity to also determine $F^{\rm deriv}_k[J]$. \step

To establish the result, we first consider the large $N$ flow \eqref{eq:largeNflowGammaBar}, \eqref{eq:Q}  with initial condition \eqref{eq:Initial} to ensure that the integrated flow has a solution of the form \eqref{eq:GammaExactSol}. By definition, the effective potenial is obtained by evaluating the effective action for constant fields. The crucial point is that the large $N$ flow \eqref{eq:largeNflowGammaBar},  \eqref{eq:Q}   only involves first derivatives $\delta \GammaInt / \delta J$, and no second derivatives $\delta^2 \GammaInt / \delta J\delta J$. Together with the result that $\GammaInt$ is a functional of fermion bilinears, 
it follows  immediately  that only interaction terms without derivatives acting on $J$ can contribute to the flow and survive the projection onto constant fields. For example, any interaction term of the form
\be\label{eq:largeNderivZbar}
\GammaInt \supset \int_x f ^A_k( J ( x ) ) \, \partial^2 J^A ( x )
\ee
will induce terms  proportional to $\partial_\mu J$ within the flow \eqref{eq:largeNflowGammaBar}, \eqref{eq:Q} after taking a $J$-derivative. These terms, however, vanish  in the constant field limit, and cannot contribute to the flow of the effective potential. If the functions $f^A$ in \eqref{eq:largeNderivZbar} are constants, the term becomes a total derivative and gives no contribution to the flow. Therefore, provided the effective action is of the form \eqref{eq:GammaExactSol}, the large $N$ flow for the effective potential decouples from the flow of higher derivative interactions, and can be solved exactly without the need to determine $F_k^{\rm deriv}[J]$.\step

Finally, we contrast  \eqref{eq:largeNderivZbar} 
with fermionic theories which include derivative interactions that cannot be expressed as functionals of the fermion bilinears \eqref{eq:JAdefPsi}. For example, the presence of derivative interactions
of the form 
\be\label{eq:morederiv}
\GammaInt \supset \int_x f_k ( J ( x ) ) \mkern4mu \chi_a ( x ) \muspace \Gamma_{ab}^\mu \muspace \partial_\mu \chi_b ( x )
\ee
implies that the interaction part of the quantum effective action, even at large $N$, is no longer a functional of fermion bilinears, \eqref{eq:GammaExactSol}, and the decomposition~\eqref{eq:d2GammaBar} is no longer applicable. Consequently, the large $N$ flow  receives contributions from derivative interactions which can survive the constant field limit, and feed into the flow of the local potential. Further, the functional flow will also generate higher derivative terms of the type \eqref{eq:morederiv},  and others, even at large $N$. In all these cases, which include generic fermionic theories at finite $N$ as well as large $N$ theories whose derivative interactions   cannot be written as functionals of fermion bilinears, we conclude that the LPA flow is not exact, and the quantum effective action will involve all types of higher derivative interactions under the RG flow.

\subsection{Flows for General Local Potentials}
\label{sec:localPotentials}

Next, we derive an explicit formula for the RG flow of general pointlike interactions at large $N$. This is achieved by passing to a momentum-space representation of the flow \eqref{eq:largeNflowGammaBar}, which allows us to evaluate the functional trace for constant field configurations. 
We  demand that the wilsonian regulator function  $R_k$ carries the same matrix structure as the kinetic term 
and write it as
\be
R_k ( q ) = i q_\mu \mkern1mu \Gamma^\mu \, r \Big( \tfrac{q^2}{k^2} \Big) \,.
\ee
This choice ensures that the cutoff term added to the action, $\psib_i \cdot R_k \cdot \psi_i$, respects the Lorentz and flavour symmetries of the free theory. The scalar function $r$ parametrises the specific shape of the cutoff. It obeys $r ( q^2/k^2 ) \to 0 $ for $k^2/q^2\to 0 $, and $r ( q^2/k^2 ) > 0$ for $q^2/k^2 \to 0$, but can be chosen freely elsewise. 
Optimised choices of cutoffs are available \cite{Litim:2000ci,Litim:2001up,Litim:2001fd,Litim:2002cf}, which aim at enhancing the stability and convergence of approximations, and permit simple analytical forms for the flow. All of the general arguments presented in this work are valid for any regulator, even though for concrete examples below, we will take specific choices for $r$.\step

With this parametrisation, the regularised free propagator, which is the matrix inverse of $\Delta^{-1}_k ( q ) = i q_\mu \mkern1mu \Gamma^\mu + R_k ( q )$, can be written in the form
\be\label{eq:Deltakq}
\Delta_k ( q ) = -i\frac{q_\mu}{q^2}\frac{ \, \Omega \mkern2mu \Gamma^\mu \mkern2mu \Omega}{1+r( q^2 / k^2)} \, ,
\ee
which is easily verified using the Clifford algebra relation~\eqref{eq:GammaClifford}. 
Inserting the propagator \eqref{eq:Deltakq} into \eqref{eq:largeNflowGammaBar} and taking the limit of constant fields, we find the  flow  for a general  interaction potential $V_k$,
\be\label{eq:LPAmain}
\partial_t V_k =  -\Nf \int \d K ( q ) \, \tr G_k ( q ) \, ,
\ee
where $G_k$ is the matrix inverse of
\be\label{eq:GinvLPA}
G_k^{-1} ( q ) = \mathds{1}_{d_\gamma} + \bigg( \frac{\slashed{q}}{q^2(1+r)}\, \sum_A \gamma^{(A)} \frac{\partial V_k}{\partial J^A} \bigg)^2 \, ,
\ee
and the trace is now only over the Dirac matrix structure. The flavour structure has been traced over, resulting in an explicit factor $\Nf$ in \eqref{eq:LPAmain}. We  also found it convenient to introduce the dressed integration measure
\be\label{eq:dK}
\d K ( q ) = \frac{\partial_t r(q^2/k^2)}{1+r(q^2/k^2)} \frac{\d^d q}{(2\pi)^d} \,.
\ee
Besides the ordinary momentum integration, it inherits the  factor ${\partial_t r}/{(1+r)} \ge 0$ from  the momentum cutoff,  whose effect it is to suppress contributions from loop momenta $q^2\gg k^2$. 
As such, the momentum cutoff effectively narrows  down the integration domain in $\d K$ to the vicinity of $q^2\lesssim k^2$. 
In the next section, we investigate  the functional  flow \eqref{eq:LPAmain} in more depth, and for various types of fermion interactions.

\section{\bf Applications}
\label{sec:applications}

In this section, we apply our method to find functional flows and conformal fixed points for different  types of large $N$ fermionic field theories in various dimensions, covering  scalar, pseudoscalar, vector,  axial-vector, and higher derivative interactions. We also discuss aspects of fermion mass generation.

\subsection{Scalar and Pseudo-Scalar Interactions}
\label{sec:SandP}

We start by considering theories with scalar and/or pseudo-scalar fermionic self-interactions. 
These theories are of the Gross-Neveu type  where interactions are built from the hermitian scalar and pseudo-scalar bilinears\footnote{Our conventions for   spinor bilinears in Euclidean  signature are detailed in  App.~\ref{sec:Dirac}.}
\be\label{eq:SandP}
S = \psib_i \psi_i \, , \quad \PS = i\psib_i \gamma^5 \psi_i \, .
\ee
Then the effective potential takes the form $V_k=V_k ( S, \PS )$ and the sum in \eqref{eq:GinvLPA} runs over the scalar and pseudo-scalar terms only.\footnote{Here we use four-dimensional terminology. In three dimensions, one can take a reducible representation of the Clifford algebra, with four-component Dirac fermions, and $\gamma^5$ directly borrowed from 4d. In this case, however, $S$ in \eqref{eq:SandP} transforms as a pseudo-scalar, and $\PS$ as a scalar, under parity transformations $x = ( x^0, x^1, x^2 ) \mapsto x' = ( x^0, -x^1, x^2 )$ with $\psi_i ( x ) \mapsto \gamma^1 \psi_i ( x' )$. In addition, one can also form the bilinears $\psib_i \gamma^3 \psi_i$ and $\psib_i i \gamma^3 \gamma^5 \psi_i$ which transform as scalar and pseudo-scalar, respectively.}
Expanding the square in \eqref{eq:GinvLPA}, one finds that $G_k^{-1}$ is diagonal in Dirac indices. After inverting and tracing, we find
\be\label{eq:SPflow}
\partial_t V_k = -d_\gamma \Nf \int \d K ( q ) \, \frac{q^2 (1+r)^2}{ q^2 (1+r)^2 + (\partial_S V_k)^2 +( \partial_{\PS} \mkern-1mu V_k)^2} \, .
\ee
The flow \eqref{eq:SPflow} can be used to extract the RG equations for all point-like polynomial 
interactions $\propto S^n S_5^m$ by projection. 
Projecting onto the scalar and pseudo-scalar four-fermion (4F) interactions,
\be
V_k ( S, S_5 ) = \ldots + \frac12 \frac{\la{ \ }}{k^{d-2}} \, S^2 + \frac12 \frac{\la{5}}{k^{d-2}} \, S_5^2 + \ldots \, ,
\ee
and also noting that the  mass terms $\propto S$, $S_5$ and  the coupling $\propto S S_5$ are  technically natural and can be consistently set to zero, we obtain the exact large-$N$ beta functions as
\begin{align}\label{eq:SP4F}
\partial_t \la{ \ } &= \left( d - 2 \right) \la{ \ } + 2 d_\gamma \Nf \, C_d [ r ] \, \la{ \ }^2 \, ,
\end{align}
and the same RG flow for $\lambda\leftrightarrow \lambda_5$. For either of these, the first term reflects the canonical mass dimension of the coupling while the second term is due to fluctuations. The coefficient $C_d [ r ]>0$ is a remnant of the operator trace in \eqref{eq:SPflow},
\be\label{eq:Cd}
C_d [ r ] = -\Omega_d \int_0^\infty \d y \, \frac{y^{d/2 - 1} \, r' ( y )}{( 1 + r ( y ) )^3} \, ,
\ee
and  encapsulates the scheme dependence due to the shape of the regulator function, while the factor $\Omega_d^{-1} = 2^{d-1} \mkern1mu \pi^{d/2} \, \Gamma ( d / 2 )$ accounts for the angular integration. 
Using an optimised regulator  \cite{Litim:2000ci,Litim:2001up,Litim:2001fd,Litim:2002cf} with shape function 
\be\label{eq:ropt}
r ( y ) = \left( {1}/{\sqrt{y}} - 1 \right) \theta ( 1 - y ) \, ,
\ee
we find $C_d [ r_{\rm opt} ] = \Omega_d / d$. 
In two dimensions, where four-fermion theories are asymptotically free and perturbatively renormalisable, their beta functions are universal with $C_2 [ r ] = \frac{1}{4 \pi}$ for any $r$, in agreement with \cite{Gross:1974jv}. 
Close to the free  fixed point, the 4F couplings scale with their classical mass dimension   $\vartheta_{\rm IR}=d-2$. For dimensions $d>2$, however, the result \eqref{eq:SP4F} implies that the canonically irrelevant 4F couplings achieve interacting UV fixed points 
\begin{equation}\label{eq:4Fscalar}
\lambda_*=\frac{2-d}{2 d_\gamma \Nf \, C_d}=\lambda_{5,*}\,, 
\end{equation}
which turns either of them  into relevant operators quantum-mechanically, with universal scaling exponent $\vartheta\equiv\partial( \partial_tx)/\partial x|_*$ for $x=\lambda,\lambda_5$, giving $ \vartheta_{\rm UV}=2-d$ for either of them at their respective  UV fixed points. Also, the  fixed points merge with the free one in the limit $d\to 2$. This type of study can be straightforwardly extended  to extract fixed points for any  pointlike polynomial interactions of the form $\lambda_{m,n}S^{m} S_5^{n}$, and  for general  regulator. \step

Conversely, the functional flow \eqref{eq:SPflow} can also be integrated {\it exactly}, without the need for an  expansion in interaction monomials. 
To that end, we employ the regulator \eqref{eq:ropt} 
so that the momentum integration in \eqref{eq:SPflow} can be evaluated analytically. Introducing the dimensionless potential $v ( \sigma, \ps ) = k^{-d} V_k ( S, \PS )$, depending on dimensionless fields $\sigma = k^{1-d} S$, $\ps = k^{1-d} \PS$, we find the large-$N$ flow for a theory with the most general local scalar and pseudo-scalar  fermionic interactions,
\be\label{eq:GNtypeLPA}
\partial_t v = - d \mkern2mu v + \left( d - 1 \right) \left( \sigma \partial_\sigma v + \ps \partial_{\ps} v \right) - \frac{1}{1 + ( \partial_\sigma v )^2 + ( \partial_{\ps} v )^2} \,.
\ee
To achieve this simple form, we have  also rescaled the fields and the potential by the number of degrees of freedom $2 d_\gamma\Nf$ and by a factor $\Omega_d / d$. 
This implies that polynomial couplings are now measured in units of perturbative loop factors and powers of $N$, in line with naive dimensional analysis.
Let us briefly discuss the significance of \eqref{eq:GNtypeLPA} for a few special cases: 

\begin{itemize}
\item[(a)] {\it Scalar Gross-Neveu Theory}\\ 
In the case where the interaction potential only  depends on $S$, 
we have that  $v=v(\sigma)$ and the flow \eqref{eq:GNtypeLPA} reduces to $\partial_t v = - d \mkern2mu v + \left( d - 1 \right) \sigma v'  - 1/(1 + v'{}^2)$ \cite{Jakovac:2013jua,Cresswell-Hogg:2022lgg,Cresswell-Hogg:2022lez} (see also 
\cite{Aoki:2014ola,Jakovac:2014lqa}). The theory possesses a discrete chiral symmetry under $\psi_i \mapsto \gamma^5 \psi_i$ provided $v$ is even under $\sigma \mapsto -\sigma$.  In terms of a hypergeometric function, the integrated flow reads  \cite{Cresswell-Hogg:2022lgg,Cresswell-Hogg:2022lez}
\be
\label{eq:analyticsolF}
\s012(d-2) \,\sigma \cdot (v')^{1-d} -
{}_2 F_1\left({\small \begin{matrix}2,1  - {d}/{2} \\2 - {d}/{2}\end{matrix}}\middle|-v'{}^2 \right)\cdot (v')^{2-d}
= G (v' e^t)\,.
\ee
The function $G$ is determined by the initial conditions $v'(\sigma)|_{k=\Lambda}$ at a reference scale $k = \Lambda$. In three dimensions, the global solutions \eqref{eq:analyticsolF} identify   interacting UV fixed points and universal scaling dimensions  in settings with \cite{Aoki:2014ola,Jakovac:2014lqa} and without \cite{Cresswell-Hogg:2022lgg,Cresswell-Hogg:2022lez} fundamental chiral symmetry. It demonstrates that the 4F fixed point \eqref{eq:SP4F} extends to a UV-complete, global fixed point for all fields whereby 6F interactions become exactly marginal. The corresponding conformal manifold terminates with the spontaneous breaking of scale symmetry and hyperscaling relations, and the appearance of a massless dilaton in the spectrum. Away from fixed points, the solution  describes UV-IR connecting trajectories, chiral symmetry breaking, and the dynamical generation of mass \cite{Cresswell-Hogg:2022lgg,Cresswell-Hogg:2022lez}. 

\item[(b)] {\it Pseudo-scalar Gross-Neveu Theory} \\
In the case where the theory only contains pseudo-scalar interactions, corresponding to powers or functions of $\PS$, 
we have that  $v=v(\ps)$ and the flow \eqref{eq:GNtypeLPA} reduces to 
\be
\partial_t v = - d \mkern2mu v + \left( d - 1 \right) \ps v'  - 1/(1 + v'{}^2)\,.
\ee
It  is identical to the scalar GN flow after substituting $\ps\to \sigma $ and $v(\ps) \to v(\sigma)$, meaning that the integrated flow for the pseudo-scalar GN theory can be read off  from  \eqref{eq:analyticsolF}. 

\item[(c)] {\it Theories with ${U(1) \times U(1)}$ Chiral Symmetry.} \\
Provided that the interaction potential depends only on the combination $S^2 + \PS^2$ and powers thereof, the resulting theory has a global $U ( 1 )_L \times U ( 1 )_R= U(1)_V \times U(1)_D$ invariance. In 4d, this is a chiral symmetry of the same type as in the one-flavour Nambu--Jona-Lasinio (NJL) model \cite{Nambu:1961tp,Nambu:1961fr}. In $d$ dimensions, the flow \eqref{eq:GNtypeLPA}  reduces to
\be\label{eq:SPcontinuousLPA}
\partial_t v = - d \mkern2mu v + 2\left( d - 1 \right) z \mkern1mu \partial_z v - \frac{1}{1 + 2 z \mkern1mu ( \partial_z v )^2} \, ,
\ee
where we have introduced $z = \frac12 \left( \sigma^2 + \ps^2 \right)$ and $v \equiv v ( z )$. 
If both $\sigma$ and $\sigma_5$ are real, we may define $\sigma_+ = \sqrt{2 z}$ and substitute $v ( z )$ by $v(\sigma_+)$. The flow \eqref{eq:SPcontinuousLPA} then takes the  same form as the flow for scalar  GN theories,   and the large $N$ exact solution to \eqref{eq:SPcontinuousLPA} is  obtained by performing the corresponding substitution in \eqref{eq:analyticsolF}. 
\end{itemize}

\subsection{Vector and Axial-Vector Interactions}\label{sec:VandA}

Next, we consider fermionic quantum field theories with vector interactions, whose local potentials $V_k=V_k ( J, J_5 )$ depend on the vector and axial-vector currents
\be
\label{eq:VandA}
J^\mu = \psib_i \gamma^\mu \psi_i \, , \quad J_5^\mu = \psib_i \gamma^\mu \gamma^5 \psi_i \,.
\ee
These types of theories display a global $U(\Nf)_L \times U(\Nf)_R$ invariance, such as in $\Nf$-flavour versions of the Nambu--Jona-Lasinio (NJL) model \cite{Nambu:1961tp,Nambu:1961fr}, and are closely related to low-energy effective models of quantum chromodynamics \cite{Braun:2011pp} or composite Higgs theories \cite{Bardeen:1989ds}. They  further reduce to models with Thirring-type interactions \cite{Thirring:1958in} provided that $V_k$ depends only on the vector current $J^\mu$. \step

To find the LPA flow \eqref{eq:LPAmain} for these theories, we note that the sum in \eqref{eq:GinvLPA} runs over the vector and axial-vector terms only. Then, after performing the algebraic operations prescribed in \eqref{eq:LPAmain}, and also using \eqref{eq:dK}, we find the exact flow for the  local potential interactions as
\be\label{eq:NJLflow}
\partial_t V_k = -\frac12 {d_\gamma \Nf}\sum_{s = \pm} \int \d K   \frac{1 - \frac{W_s^2}{q^2(1+r)^2} +  \frac{2 (q \cdot W_s )^2}{q^4(1+r)^2}}{\left(1 -  \frac{W_s^2}{q^2(1+r)^2} \right)^2 + \frac{4 (q \cdot W_s )^2}{q^4(1+r)^2}} \, ,
\ee
where
\be
W_\pm^\mu =  \frac{\partial V_k}{\partial J_\mu} \pm \frac{\partial V_k}{\partial J_{5 \mu}}  \, .
\ee
In comparison with the flow for theories with scalar interactions \eqref{eq:SPflow}, we notice that the operator trace for vector interactions now also depends on angles due to terms involving $(q\cdot W_s)^2$.\step

Introducing dimensionless 4F couplings $g$ and $g_5$ to denote the squared vector and axial-vector interactions,
\be
V_k ( J, J_5 ) = \ldots + \frac12 \, \frac{g}{k^{d-2}} \, J^2 + \frac12 \, \frac{g_5}{k^{d-2}} \, J_5^2 + \ldots\,,
\ee
and projecting the flow \eqref{eq:NJLflow} onto these couplings via the same procedure as in the scalar case, we find their exact large $N$ beta functions in general dimension $d$ as
\begin{align}\label{eq:beta4FVA}
\partial_t g_{{}\ } &= \left( d - 2 \right) g_{{}\ } + 2 d_\gamma \Nf \left( \tfrac{2}{d} -1\right) C_d [ r ] \, g^2 \,,
\end{align}
and the same RG flow for $g\leftrightarrow g_5$.\footnote{In four dimensions, the flow of  4F interactions in the form $\lambda_\pm(J^2 \pm J_5^2)$ has been studied in  \cite{Braun:2011pp}. Accounting for differences in conventions for euclidean Dirac matrices and definitions of the conjugate field $\psib$, our general result \eqref{eq:beta4FVA}  for $d=4$ maps exactly onto Eqs.~(117)~and~(118) of~\cite{Braun:2011pp}, also using $\la{\pm} = - \tfrac12 \left( g \mp g_5 \right)$.} In comparison with the result for scalar and pseudoscalar 4F interactions  \eqref{eq:SP4F}, the sole but noteworthy difference  is that the scheme-dependent coefficient $C_d [ r ]$ is now replaced by $(\tfrac2d-1)C_d [ r ]$.  The reason for this relates to the vector nature of  interactions, which is responsible for an angular dependence under the operator trace $\propto(q\cdot J)^2$ and $\propto(q\cdot J_5)^2$ in \eqref{eq:NJLflow}. In consequence, and after projection onto the 4F couplings, an additional factor $\cos 2 \vartheta$ alters the angular integration, giving  $(\tfrac{2}{d} - 1)\Omega_d$ rather than $\Omega_d$. 
\step

In two dimensions, where 4F theories are perturbatively renormalisable, the vector nature of 4F interactions has an important implication. In fact,  the result \eqref{eq:beta4FVA} states that no quantum corrections arise to the running of the point-like vector and axial-vector 4F interactions at large $N$, much like in ${\cal N}=4$ super-Yang-Mills theory in four dimensions. This is very different from what happens for scalar-type interactions (see Sec.~\ref{sec:SandP}), where the asymptotically free couplings  run even at large $N$ to trigger dynamical mass generation and chiral symmetry breaking \cite{Gross:1974jv}. In theories with vector interactions, instead,  the classically marginal couplings $g$ and $g_5$  remain exactly marginal interactions even quantum-mechanically, and their values parametrise  lines of fixed points at large $N$. Since either of these are  marginal deformations of the free theory, they imply a two-dimensional conformal manifold with  classical scaling dimensions. 
In this context, it is worth noting that the exact marginality of the $g J^2 $  interaction in two dimensions has previously been noticed by Dashen and Frishman \cite{Dashen:1974hp} based on an integrable Thirring model involving abelian and non-abelian vector currents.\footnote{Specifically, Ref.~\cite{Dashen:1974hp} studies two 4F interactions, $g_B J^2$ and $g_v (\psib \gamma^\mu \lambda^a \psi) (\psib \gamma_\mu \lambda^a \psi)$, where $\lambda^a$ are generators of $SU(N)$ in the fundamental representation, finding that $g_B$ is exactly marginal and $g_v$ is asymptotically free. The latter of the two interactions can be expressed in terms of flavour-singlet bilinears using the completeness relation for $SU(N)$ generators and Fierz identities (see App.~\ref{app:2nFinteractions}), and is equivalent to a linear combination of $S^2$, $S_5^2$ and $\frac{1}{N} J^2$ in the notation of this work.  Hence $g_B$ is identical to the coupling $g$ in \eqref{eq:beta4FVA} at large $N$.  Our study adds the insight that the $g_5 J_5^2$ interaction also becomes exactly marginal, and for the same underlying reasons.}
\step

Close to the free Gaussian fixed point, the 4F vector and axial-vector couplings scale with their classical mass dimensions   $\vartheta_{\rm IR}=d-2$. For dimensions $d>2$, and also recalling \eqref{eq:Cd}, the result \eqref{eq:beta4FVA} implies that the canonically irrelevant 4F couplings achieve interacting fixed points 
\begin{equation}\label{eq:4Fvector}
g_*=\frac{d}{2 d_\gamma \Nf \, C_d}= g_{5,*}\,, 
\end{equation}
which turn $\int J^2 $ and $\int J_5^2$ into relevant operators quantum-mechanically, with universal scaling dimension $\vartheta_{\rm UV}=2-d$. 
Notice that both interacting fixed points have a finite limit for $d\to 2$. However, we have shown previously that in $2d$ all coupling values are equivalent, corresponding to a line of fixed points. 
Hence, the exact marginality of interactions at $d=2$ is not captured from \eqref{eq:4Fvector} in the limit  $d\to 2$. 
This pattern of results is  also different from theories with scalar interactions, which in the limit $d\to 2$ merge with the free fixed point in $2d$, see \eqref{eq:4Fscalar}. \step

This study can  be extended  to the RG flows for higher order pointlike vector and axial-vector interactions, and  
for general  regulator shape function. Similarly, for suitable regulators \cite{Litim:2000ci,Litim:2001up,Litim:2001fd,Litim:2002cf} including \eqref{eq:ropt}, closed analytical expressions can be found for the full flow \eqref{eq:NJLflow}. Their detailed analysis is beyond the scope of this work and  will be reported elsewhere.

\subsection{Derivative Interactions}
\label{sec:higherDerivatives}

Next, we turn to derivative interactions. At large $N$,  the general form \eqref{eq:GammaExactSol} of the effective action also includes all possible interactions built from derivatives of the bilinears $J^A ( x )$. For want of example, we consider the impact of a derivative term of the type \eqref{eq:largeNderivZbar} in a Gross-Neveu theory at large $N$. 
We show that derivative interactions of this type are always generated by pointlike interactions, 
and that they take an interacting fixed point provided the potential interactions do so in the first place. 
Specifically, we consider the subspace of interactions built out of the scalar bilinear
\be
S ( x ) = \psib_i ( x ) \mkern1mu \mkern1mu \psi_i ( x ) \,.
\ee
We approximate the interaction part of the effective action as
\be
\GammaInt [ \chi ] \approx \int_x \left\{ V_k ( S ( x ) ) - \tfrac12 G_k \mkern2mu S ( x ) \partial^2 S ( x ) \right\} \, ,
\ee
expanding on the form \eqref{eq:LPAinteractions} by the inclusion of the higher derivative four-fermion interaction with coupling $G_k$. This coupling has mass-dimension $[G_k] = -d$ at the classical level and can be projected out of the action via
\be\label{eq:GkProjection}
(2\pi)^d \delta ( 0 ) \, G_k = \delta_{p^2} \! \left( \int_x \int_y e^{i p \cdot ( x - y )} \frac{\delta^2 \GammaInt}{\delta S ( x ) \delta S ( y )}\bigg\rvert_{S_0} \right) \, .
\ee
In this expression, $S_0$ stands for a constant field configuration and $\delta_{p^2}$ is an operator which extracts the coefficient of $p^2$, acting on functions of momenta as
\be\label{eq:psquaredOp}
\delta_{p^2} ( X ) = \frac{1}{2d} \frac{\partial}{\partial p_\mu} \frac{\partial}{\partial p^\mu} X ( p ) \big\rvert_{p=0} \, .
\ee
The flow for the derivative coupling $G_k$ is obtained by applying the projection  \eqref{eq:GkProjection} to the functional flow \eqref{eq:largeNflowGammaBar}. Doing so, we find
\be\label{eq:dtGkTraced}
\partial_t G_k = 2 d_\gamma \Nf \int \d K'\left\{ 2 \mkern1mu G_k \mkern1mu V''_k ( S_0 ) \,  P ( q^2 ) + V''_k ( S_0 )^2 \muspace \left[ \left( 1 + \frac{2}{d} \right)  P'(q^2) + \frac{2}{d} \, q^2 P''(q^2) \right] \right\} \, ,
\ee
where we have introduced the scalar propagator
\be\label{eq:PpropDef}
P ( q^2 ) = \frac{1+r}{q^2(1+r)^2  +V'_k ( S_0)^2} \,,
\ee
which depends on the running fermion mass $M_F\equiv V'_k ( S_0 )$. Primes on the propagator indicate differentiation with respect to the argument, e.g.~$P'(q^2)= \partial P(q^2)/\partial{q^2}$. 
We  also found it convenient to introduce the modified integration measure
\be\label{eq:dK'}
d{K'} = \frac{d^dq}{(2\pi)^d} \frac{q^4(1+r)^2 \,\partial_t r}{[q^2(1+r)^2  +V'_k ( S_0)^2]^2} \,,
\ee
which  relates to \eqref{eq:dK}  as $dK'=dK/(1+r)$ in the limit of a vanishing fermion mass.\step

Two types of vertex structures appear in the flow \eqref{eq:dtGkTraced}, proportional to $G_k V''_k$ and $(V''_k)^2$, meaning that the flow \eqref{eq:dtGkTraced} for the derivative coupling $G_k$ is driven by the derivative coupling itself,  and by the pointlike (non-derivative) 4F coupling $V''_k$. A dependence on the running fermion mass $V'_k $ only enters  through \eqref{eq:PpropDef}. 
Also, contributions proportional to $G_k^2$ do not appear in \eqref{eq:dtGkTraced}  because they come with a factor $\sim p^4$ and generate 4F interactions with four derivatives. 
We conclude from the inhomogeneous contributions $\propto  V''_k$ that derivative 4F interactions are inevitably   induced by the pointlike 4F interactions, regardless of the initial value of $G_k$, even though they will not couple back into the exact running of point-like interactions. This  result illustrates the general arguments presented in Sec.~\ref{sec:exactSols}.\step

Explicit evaluation of the momentum integration in \eqref{eq:dtGkTraced} can be achieved using the optimised regulator \eqref{eq:ropt}.\footnote{The relevant momentum traces involve products of distributions such as $\int dx\,  \delta(x)\,F[\theta(x)]=\int_0^1dz\, F(z)$ for  smooth functions $F(z)$, see \cite{Morris:1993qb}.} We find 
\be\label{eq:flowDerivLargeN}
\partial_t g_{\rm 4F}  = d\muspace g_{\rm 4F}  +  {d_\gamma \Nf}\, \frac{\Omega_d}{d} \left( \frac{4 g_{\rm 4F} }{(1+m^2)^3} 
- \frac{4-d\left( 3 + m^2 \right) }{(4-2d)(1+m^2)^4} \,\lambda_{\rm 4F} \right) \lambda_{\rm 4F} \, ,
\ee
where the dimensionless couplings $g_{\rm 4F}  = k^d G_k$,  $\lambda_{\rm 4F}  = k^{d-2} V''_k$ and $m \equiv M_F/k = V'_k/k$  correspond, respectively,  to the  two-derivative 4F coupling,  the  pointlike 4F coupling, and the fermion mass in units of the RG scale $k$. 
The factor $\Omega_d$ is the solid angle arising from angular integration. 
We remark that fluctuations are suppressed and the running of $g_{\rm 4F}$ becomes entirely classical both in the decoupling limit, where the fermion mass $M_F$ becomes large compared to the RG scale, $m\gg 1$, and in the limit where the 4F interactions become negligible $|\lambda_{\rm 4F}|\ll 1$.\step

\begin{figure}[t]
\includegraphics[width=.5\linewidth]{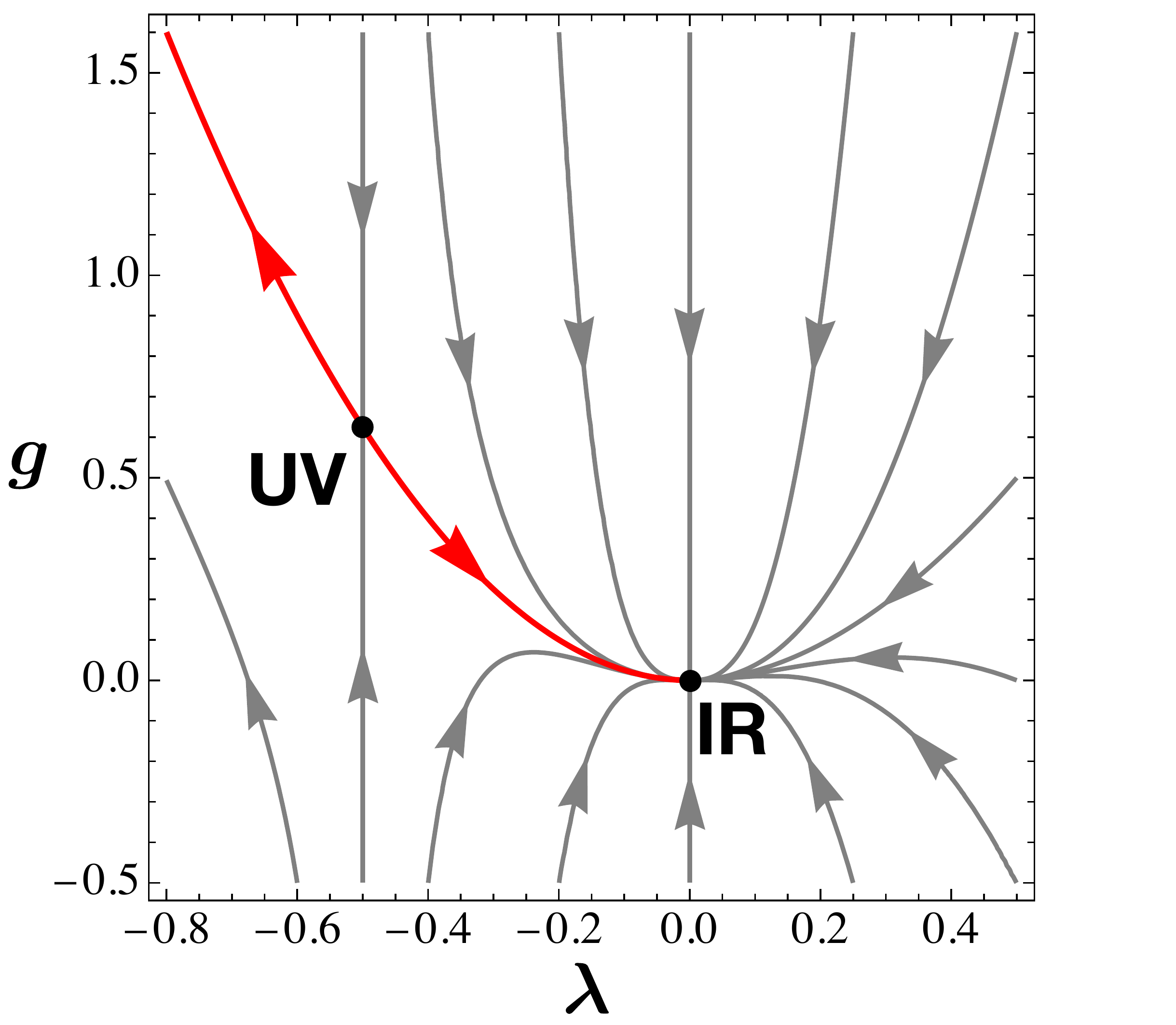}
\caption{RG flow \eqref{eq:lag_4F} in the $(\lambda_{\rm 4F} , g_{\rm 4F})$-plane at large $N$ in three dimensions, with arrows indicating the direction of flow from UV to IR. Highlighted in red are UV-complete separatrices which emanate from the interacting UV fixed point. All trajectories to the right of the UV fixed point ($\lambda_{\rm 4F}  > -1/2$) terminate at the free theory in the IR, while those to the left ($\lambda_{\rm 4F}  < -1/2$) run to a strong coupling regime with dynamical mass generation.}
\label{fig:4Fflow}
\end{figure}

Let us now study \eqref{eq:flowDerivLargeN} in the massless limit $V'_k(S_0)=0$. 
Fermion mass can be protected by demanding invariance under chiral symmetry, or, more generally, by imposing the large $N$ limit as done here (see Sec.~\ref{sec:fermionMass}).
Taking two derivatives of the flow \eqref{eq:LPAmain} with respect to the bilinear $S$, and evaluating at vanishing fields $S_0 = 0$, allows us to extract the flow for the pointlike four-fermion coupling $\lambda_{\rm 4F}$. After rescaling  the four-fermion couplings $\lambda_{\rm 4F}$ and $g_{\rm 4F}$ by the number of fermions $d_\gamma \Nf$, and  by the factor $\Omega_d / d$ which is a remnant of the operator trace \eqref{eq:Cd} with optimised regulator  \eqref{eq:ropt}, we find the coupled system of RG equations,
\begin{subequations}\label{eq:lag_4F}
\begin{align}
\partial_t \lambda_{\rm 4F}  &= \left( d - 2 + 2 \lambda_{\rm 4F}  \right) \lambda_{\rm 4F}  \, , \\
\partial_t g_{\rm 4F}  &= \left( d + 4 \lambda_{\rm 4F}  \right) g_{\rm 4F} - \frac{4 - 3d}{4 - 2d} \,\lambda_{\rm 4F}^2 \,.
\end{align}
\end{subequations}
We observe that both flows are driven by the point-like $\lambda_{\rm 4F}$  interaction. Also, the two-derivative interaction couples back to itself, but does not inform the flow of $\lambda_{\rm 4F}$, in accord with our general findings.  The system of flow equations \eqref{eq:lag_4F} can be integrated exactly. It  displays two fixed points, the free IR fixed point with $\lambda_{\rm 4F}  = g_{\rm 4F}  = 0$, and an interacting UV fixed pont with
\be\label{eq:omega}
\lambda_{\rm 4F} ^* = \frac{2-d}{2}, \quad g_{\rm 4F} ^* = \frac{ \left( d - 2 \right) \left( 3d - 4 \right)}{8 ( 4 -d )}\,.
\ee
The latter characterises the non-perturbative renormalisability  of the Gross-Neveu theory in $2<d<4$ dimensions, e.g.~\cite{Wilson:1972cf,Gawedzki:1985ed,Rosenstein:1988pt,Kikukawa:1989fw,deCalan:1991km,Gies:2010st,Braun:2010tt,Braun:2011pp,Jakovac:2014lqa,Gehring:2015vja,Dabelow:2019sty,Cresswell-Hogg:2022lgg,Cresswell-Hogg:2022lez}. 
For illustrative purposes, the flow in three dimensions is depicted in Fig.~\ref{fig:4Fflow}.\step

Linearising the flow \eqref{eq:lag_4F} around the fixed point \eqref{eq:omega}, one finds a  relevant eigenperturbation $\sim \int  S^2$ associated to the point-like 4F interaction with eigenvalue $\theta_\lambda = 2 - d<0$. The   
eigenperturbation  $\sim \int  S\partial^2 S$ associated to the two-derivative 4F interaction  is identified as irrelevant due to its positive  eigenvalue $\theta_g =  4 - d>0$. 
The latter becomes exactly marginal for $d\to 4$, where the fixed point $g^*_{\rm 4F}$ ceases to exist. 
\step

In three dimensions, our results for fixed points and scaling dimensions also agree with  previously computed higher-derivative 4F interactions by Dabelow, Gies, and Knorr \cite{Dabelow:2019sty}. 
Given that the derivative interaction with the smallest canonical mass dimension $G\int  S \partial^2 S$  already leads to an irrelevant perturbation, further interaction monomials with either more derivatives, or more powers in the fields, or both, will  lead to increasingly irrelevant  eigenperturbations   \cite{Jakovac:2014lqa,Dabelow:2019sty,Cresswell-Hogg:2022lgg,Cresswell-Hogg:2022lez}.\step

With the above results at hand, it is interesting to look at how universal scaling dimensions of operators or interaction monomials change when the theory interpolates between the interacting UV fixed point and the free IR fixed point. 
Using \eqref{eq:lag_4F}, scaling dimensions associated to the 4F interaction  ${\cal O}^{\rm 4F}_0\sim \,k^{2-d}\int S^2$ and the two-derivative 4F interaction ${\cal O}^{\rm 4F}_1\sim \,k^{-d}\int S\partial^2 S$ 
can be extracted  from the running couplings at the interacting UV fixed point $\{\theta^{\rm UV}_i,\,i=0,1\}$ and  
the  free  IR fixed point, $\{\theta^{\rm IR}_i\}$. 
We are particularly  interested in the eigenvalue shifts 
\begin{equation}\label{eq:shift}
\Delta \theta_i\equiv \theta_i\big|_{\rm UV}-\theta_i\big|_{\rm IR}
\end{equation}
 induced by  quantum fluctuations. Notice that 
 the shifts are the same as the shifts in the corresponding operator scaling dimensions.\footnote{If we denote the scaling dimension of a composite operator ${\cal O}$ by $\Delta_{\cal O}$, then the shift in the operator scaling dimension is defined as $\Delta_{\cal O}|_{\rm UV}-\Delta_{\cal O}|_{\rm IR}$. This relates to \eqref{eq:shift} as $\Delta \theta_i=\Delta_{{\cal O}_i}|_{\rm UV}-\Delta_{{\cal O}_i}|_{\rm IR}$ for ${\cal O}_i\equiv {\cal O}^{\rm 4F}_0$ or ${\cal O}^{\rm 4F}_1$.} 
Even though  eigenoperators  in the UV are (mildly) different from those in the IR, we can relate them unambiguously along UV-IR connecting trajectories. 
We  find that the fluctuation-induced eigenvalue shifts \eqref{eq:shift} for the 4F  and the two-derivative 4F  interactions $ \Delta \theta_i\equiv \Delta_{4\rm F}$ read
\be\label{eq:uni}
\Delta_{4\rm F}=4-2d\,.
\ee 
Notice that the result is  {\it universal}  in that the shifts only depend on  space-time dimensionality,  but are {\it independent} of the number of derivatives contained in the 4F interaction monomials, and valid in the range of dimensions $2\le d<4$. \step

Interestingly, in three dimensions,  the $s$-channel momentum dependence of the four-fermion vertex at large $N$, together with the UV and IR eigenvalue spectra, have been determined in    \cite{Dabelow:2019sty}. 
When expanded in powers of derivatives,  we observe that the result  for the shifts in scaling dimensions \eqref{eq:uni}    is in fact  valid for  {\it all} higher-derivative 4F interaction monomials in the $s$-channel,  schematically  written as  ${\cal O}^{\rm 4F}_m\sim \,k^{-d-2m+2}\int S\partial^{2m}S $. The  result  further corroborates that the classical-to-quantum shifts \eqref{eq:uni} are entirely due to the 4F nature of interactions, but insensitive to the number of derivatives contained in them.\step

Finally, given that the exact eigenvalue shifts for all point-like $2n$F interactions are known \cite{Jakovac:2014lqa,Cresswell-Hogg:2022lgg,Cresswell-Hogg:2022lez},  and {\it assuming} that  the pattern \eqref{eq:uni}  observed for 4F interactions persists for general $2m$-derivative 2nF interaction monomials  (schematically  ${\cal O}^{\rm 2nF}_m\sim \,k^{-d(n-1)-2m+n}\int S\partial^{2m} S^{n-1}$), we conjecture that their shifts in eigenvalues and scaling dimensions
universally read
\be \label{eq:unin}\Delta_{2n\rm F}=2n-2d\ee  
in the large-$N$ limit. An explicit confirmation of the latter (for $n>2$)  is left for future study.

\subsection{Fermion Mass}
\label{sec:fermionMass}

Finally, we discuss aspects of fermion mass generation at large $N$. Mass terms in fermionic systems are often protected by discrete or continuous symmetries such as parity or chiral symmetry. Still, fermion mass can be generated  via strong dynamics,  leading to the dynamical breaking of a symmetry.
If fermion mass is not protected by a symmetry, mass can also be generated by fluctuations, without the need for strong coupling. \step

Recent studies in Gross-Neveu theories  revealed   that fluctuation-induced fermion masses, in the absence of chiral symmetry, are $1/N$  suppressed over the the mass generated by strong dynamics \cite{Cresswell-Hogg:2022lgg,Cresswell-Hogg:2022lez,Cresswell-Hogg:2023hdg}.
Here, we briefly discuss the fate of fluctuation-induced masses in general fermionic theories at large $N$. To that end, it is sufficient to consider the two-point functions at zero momentum, corresponding to constant fields in position space, which are conveniently extracted from the general local potential flow \eqref{eq:LPAmain}. Physical field configurations are solutions of the vacuum equations of motion, $\delta \Gamma_k / \delta \chi_a = 0$, and the only solution which is both constant in space and Poincar{\' e} invariant is the homogeneous configuration $\chi_a ( x ) \equiv 0$. Given the set of independent fermion bilinears $\psib\gamma^{(A)}\psi$, it is convenient to introduce a mass term for each of them by writing
\be\label{eq:mass}
m_\psi^A = \left. \frac{\partial V_k}{\partial J^A} \right\rvert_{J = 0} \, .
\ee
In particle physics, Lorentz symmetry dictates that only scalar mass terms $m_\psi \psib\psi$ and pseudo-scalar mass terms   $m_\psi^5 \psib i\gamma^5\psi$ are permitted. 
Further, in 3d theories with reducible four-component spinors, one finds multiple independent scalar and pseudo-scalar mass terms due to the presence of both $\gamma^3$ and $\gamma^5$.
Non-Lorentz-invariant mass terms may be of relevance as well, for example as order parameters in non-relativistic condensed matter systems including graphene \cite{Gehring:2015vja,Parthenios:2023apm}.\step

The flows for  the mass terms \eqref{eq:mass} are  obtained from the general potential flow \eqref{eq:LPAmain} by taking a $J^A$ derivative and evaluating at vanishing fields, $\partial_t m^{A}_k=\partial(\partial_t V_k)/\partial J^A|_{J=0}$. We find
\be\label{eq:massFlow}
\partial_t m_\psi^A = \Nf \int \d K ( q ) \, \left. \tr{ G_k ( q ) \, \frac{\partial G_k^{-1} ( q )}{\partial J^A} \, G_k ( q ) } \right\rvert_{J = 0} \, .
\ee
Computing the derivative inside the trace yields
\be\label{eq:massNatural}
\left. \frac{\partial G_k^{-1} ( q )}{\partial J^A} \right\rvert_{J=0} \propto 
\sum_{B,C} \left\{ \slashed{q} \, \gamma^{(B)} , \, \slashed{q} \, \gamma^{(C)} \right\} V^{(2)}_{AB} \mkern4mu m_\psi^C \, ,
\ee
which depends only on the masses themselves and the various four-fermion couplings $V^{(2)}_{AB}$. 
We conclude that the running of mass terms takes the  form
\be\label{eq:dtmA}
\partial_t m_\psi^A = \sum_B D^{AB} \,m_\psi^B
\ee
where the dimensionless matrix $D$ depends on the masses and the various 4F couplings $V^{(2)}_{AB}$. The mass dependence  enters explicitly via \eqref{eq:massNatural}, and implicitly through the propagators $G(q)$. \step

The absence of inhomogeneous terms in \eqref{eq:dtmA}, i.e.~terms not proportional to masses, implies that none of the mass terms can be switched on by interactions alone. 
 In the presence of symmetries that forbid  fermion mass, the standard picture of a technically natural fermion mass in the sense of 't~Hooft~\cite{tHooft:1979rat} applies, including that the absence of mass enhances a symmetry. 
Notice however  that  no assumptions about symmetries have entered the derivation of \eqref{eq:dtmA}.  Then, in settings where mass is {\it not} protected by a symmetry,  fermion mass remains technically natural, though  with the large $N$ caveat that a vanishing mass, in this case,  does not  enhance a  symmetry.\step

Further, we have emphasised previously that at large $N$, the local potential flows are exact for {\it any} subset of bilinears $J$. Here, this  implies that the form \eqref{eq:dtmA} is valid for any subset of masses. 
Consequently, for any and all subsets of fermion mass terms \eqref{eq:mass}, the generation of  mass by fluctuations  is suppressed as $1/N$  at large $N$, and   protected at infinite $N$, even if mass is not protected by a  
symmetry. Interestingly, the pattern of results  entails    that the {\it dynamical}  generation of fermion mass   proceeds  through a continuous phase transition, irrespective of symmetry, which at finite $N$ turns into a smooth cross-over  provided that  mass is not protected by a symmetry, see  \cite{Cresswell-Hogg:2024pxd}. We conclude that the pattern of fermion mass generation uncovered in   \cite{Cresswell-Hogg:2022lgg,Cresswell-Hogg:2022lez,Cresswell-Hogg:2023hdg,Cresswell-Hogg:2024pxd}  for fermionic theories with scalar-type interactions  is  {\it genuine} in that it  holds  true for    large-$N$ theories with  the most general type of $U(N)$ symmetric  interactions.

\section{\bf Discussion and Conclusions}
\label{sec:conclusion}

We have put forward a comprehensive study of fermionic quantum field theories,  from first principles and in general dimensions, by combining functional renormalisation with  a large $N$ limit, where $N$ relates to the number of fermion flavours. 
The virtue of our setup is that it admits 
exact solutions for  quantum effective actions  in the form \eqref{eq:GammaExactSol}, with interactions built exclusively out of a set of flavour-singlet fermion bilinears \eqref{eq:JAdefPsi} and their derivatives. 
Fierz ambiguities are $1/N$  suppressed, which ensures closure of RG flows for theories whose interaction functionals depend on {\it any} subset of fermion bilinears  \eqref{eq:JAdefPsi}. Our setup benefits from  choices for the fermion basis \eqref{eq:chiDef}, \eqref{kinetic} and  the set of independent flavour-singlet fermion bilinears, and  exploits the underlying symplectic and Clifford algebra structures, and global symmetries. As such, it provides a practical starting point for the study of strongly-coupled fermionic theories  at large $N$ and beyond.\step 

We have further provided the  renormalisation group  flows  for    fermionic theories   in the local potential approximation with the most general microscopic interactions   \eqref{eq:Initial},  see \eqref{eq:LPAmain} with \eqref{eq:GinvLPA}, \eqref{eq:dK}, and for any type of wilsonian momentum cutoff. Notable features  at large $N$ are that fermion anomalous dimensions vanish ($\eta_\psi= 0$), that higher derivative interactions decouple from local potential interactions, and that the local  potential flow becomes closed, exact, and exactly solvable. Hence, masses and $2n$-point functions at vanishing  momenta can be determined exactly including at strong coupling,  and up to corrections suppressed as $1/N$. Higher derivative interactions also arise, inevitably sourced by point-like interactions. Notice that fermions may develop non-trivial  anomalous dimensions $(\eta_\psi\neq 0)$, which follows from the general structure of the Hessian, \eqref{eq:d2GammaBar}. At infinite $N$, however, a necessary and sufficient condition for $\eta_\psi\neq 0$  is that microscopic interactions are {\it not} of the form  \eqref{eq:Initial}. Then, and only then, terms such as \eqref{eq:morederiv} are generated by fluctuations that cannot be reduced to interactions of fermion bilinears, and the  form \eqref{eq:GammaExactSol} remains unachievable for any scale.  For these types of theories,  the local potential approximation is never exact. \step 

To illustrate our techniques,  we have identified  conformal critical points in various fermionic theories. New results include functional flows for Gross-Neveu theories with scalar and pseudo-scalar interactions, \eqref{eq:SPflow}, and theories with  or without discrete or continuous chiral symmetry,   also covering  solutions for  local potential interactions \eqref{eq:analyticsolF} and their  fixed points and critical exponents, e.g.~\eqref{eq:4Fscalar}. We have equally  provided  functional flows for  Nambu-Jona-Lasinio-type theories  with vector- and axial-vector interactions   \eqref{eq:NJLflow}, and the fixed points of their four-fermion couplings~\eqref{eq:4Fvector}. Interestingly, in two dimensions, theories  with vector or axial-vector interactions  can develop lines of conformal fixed points, much like in maximally supersymmetric Yang-Mills theory, and quite different from asymptotic freedom  and dynamical mass generation as displayed by theories with scalar or pseudo-scalar interactions  \cite{Gross:1974jv}. The culprit for this difference is  the vector nature of interactions which introduces an angular dependence in operator traces that make the large-$N$ leading quantum corrections vanish, see  \eqref{eq:SP4F} vs \eqref{eq:beta4FVA},   opening up  entire conformal manifolds worthy of further study.\step

Next, we comment on the role of higher derivative interactions. In derivative expansions, higher derivative interactions are  radiatively induced along functional flows \eqref{eq:functionalFlowMain}, \eqref{eq:flowGammaBar}, even if they are absent microscopically. Their  back-coupling is $1/N$ suppressed, however, and implies decoupling at large $N$ as long as interaction functionals take the  form \eqref{eq:GammaExactSol}.  At critical points, they take fixed points  by themselves (Fig.~\ref{fig:4Fflow}), but elsewise remain irrelevant operators quantum-mechanically. Also, the large-$N$ quantum-induced shifts of  operator scaling dimensions at critical points are found to be {\it universal} in that they only depend on  the space-time dimensionality and the number of fermions  contained in  interaction monomials,  but not on the number of derivatives, see \eqref{eq:shift}, \eqref{eq:uni}, and \eqref{eq:unin}. It will be interesting to confirm these findings with other methodologies, e.g.~perturbation theory, the conformal bootstrap, or the lattice.\step

From the viewpoint of mass generation, we recall that fermion mass is  oftentimes protected by a discrete or continuous global symmetry such as parity or chiral symmetry, and  technically natural in the sense of `t~Hooft~\cite{tHooft:1979rat}. Mass can  still be generated  dynamically,  with or without the breaking of a symmetry. If fermion mass is {\it not} protected by a symmetry, the absence of inhomogeneousn terms in \eqref{eq:dtmA} demonstrates that  radiatively-induced  masses are $1/N$ suppressed. This holds true for  general fermionic theories with  $U(N)$ symmetric interactions \eqref{eq:Initial}, and for  any and all subsets of  fermion mass terms \eqref{eq:mass}, in accord with  recent findings in Gross-Neveu theories, see \cite{Cresswell-Hogg:2022lgg,Cresswell-Hogg:2022lez,Cresswell-Hogg:2023hdg}. We conclude  that  fermion masses cannot be produced  by radiative corrections at large $N$, irrespective of symmetry and interactions. This also  implies  that the {\it dynamical}  generation of mass in large-$N$ theories proceeds through a continuous quantum phase transition \cite{Cresswell-Hogg:2024pxd}.\step

In future work, it will  be interesting to systematically exploit our setup   to investigate   critical points in fermionic theories and models of particle physics, both at large-$N$ and beyond. 
It will    also be    instructive to establish  links with  composite field formulations using  bosonisation ideas~\cite{Gies:2001nw,Pawlowski:2005xe} 
or  functional dualities   \cite{Weinberg:1997rv,Cresswell-Hogg:2023hdg}.
Other promising directions include the search for  theories with spontaneously broken scale symmetry which necessitate  critical points with exact moduli spaces of degenerate vacua \cite{Cresswell-Hogg:2022lez,Cresswell-Hogg:2025kvr}, 
and the extraction of conformal data  from the renormalisation group \cite{Cardy:1996xt} to complement the  bootstrap programme \cite{Chen:1992ig,Karateev:2019pvw,Erramilli:2022kgp}.\step\step

{\bf Acknowledgements.} This work has been supported by the Science and Technology Facilities Council (STFC) under the Consolidated Grant No. T/X000796/1,  and  by a CERN Associateship.

\appendix
\renewcommand{\thesection}{{\bf \Alph{section}}}
\renewcommand{\theequation}{\Alph{section}\arabic{equation}}

\section{\bf Completeness of Pointlike Interaction Basis}
\label{app:2nFinteractions}

Our basis for pointlike interactions in the effective action is the set of flavour-singlet bilinears $J^A = \psib_i \gamma^{(A)} \psi^i$. This basis is complete, in the sense that any $U(\Nf)$-invariant $2n$-fermion interaction term without derivatives can be reduced by repeated application of Fierz identities to a form depending only on the singlet bilinears $J^A$. In this appendix we provide a proof of this statement.\step

{\bf Fermion self interactions.} \
With $\Nf$ flavours of Dirac fermion transforming in the fundamental representation, the most general $U(\Nf)$ symmetric, local $2n$-fermion interaction term without derivatives is of the form
\be\label{eq:general2nF}
c_{A_1 \cdots A_{n}} \, f^{i_1 \cdots i_{n}}_{j_1 \cdots j_{n}} \, ( \psib_{i_1} \gamma^{(A_1)} \psi^{j_1} ) \, ( \psib_{i_2} \gamma^{(A_2)} \psi^{j_2} ) \, \cdots \, ( \psib_{i_{n}} \gamma^{(A_n)} \psi^{j_{n}} ) \, ,
\ee
where Dirac indices are contracted within each set of parentheses and summation over repeated indices is implied. We distinguish between fundamental and antifundamental flavour indices using upper and lower index placement. Lorentz invariance constrains the coefficients $c_{A \cdots}$, but this will not play a role in the analysis. The global flavour symmetry constrains $f^{i_1 \cdots i_{n}}_{j_1 \cdots j_{n}}$ to be a $U(\Nf)$ invariant tensor, the basic invariant tensor being the Kronecker delta $\delta^i_j$. \step

{\bf Invariant tensors.} \
The first consideration is that of invariant tensors. In $U(\Nf)$, all invariant tensors can be built by composing Kronecker deltas using tensor products and linear combinations \cite{Cvitanovic:2008zz}. For our purposes, this means that we only need to consider flavour tensors in \eqref{eq:general2nF} which can be written as the tensor product of $n$ Kronecker deltas. 
We emphasise at this point that fermion bilinears involving flavour generators do not need to be considered separately. They are automatically included in the analysis because adjoint indices must be contracted so as to create invariant tensors in the fundamental representation. The simplest example is a 4F term $(\psib T^a \psi) (\psib T^a \psi)$, $a$ being the adjoint index, which is of the form \eqref{eq:general2nF} with $f^{ik}_{jl} = (T^a)^i_j (T^a)^k_l$. The latter reduces to Kronecker deltas by the completeness relation for the generators.\step

In $SU(\Nf)$, one also has the antisymmetric epsilon tensors $\epsilon_{i_1 \cdots i_{\Nf}}$ and $\epsilon^{i_1 \cdots i_{\Nf}}$ as basic invariants. 
Although we are concerned only with global $U(\Nf)$ symmetry in the body of this work, the analysis of this section equally covers these $SU(\Nf)$ invariants, which can arise, for instance, from instanton interactions in effective descriptions of quantum chromodynamics \cite{Schafer:1996wv}. The reason for this is that the Dirac fields and their conjugates have to appear in pairs, meaning an equal number of upper and lower indices on the flavour tensor $f$. Thus any term involving one epsilon tensor, say with upper indices, must have a second with lower indices. The resulting tensor product of epsilon tensors defines a generalised Kronecker delta,
\be
\epsilon^{i_1 \cdots i_{\Nf}} \epsilon_{j_1 \cdots j_{\Nf}} = \delta^{i_1 \cdots i_{\Nf}}_{j_1 \cdots j_{\Nf}} \, ,
\ee
which is itself built from tensor products of $\delta_i^j$ by summing over signed permutations of indices. Contracting any pair of indices on the generalised delta gives a generalised delta of lower rank.\step

Therefore, in all of the above cases, the flavour tensor $f_{i_1 \cdots i_{n}}^{j_1 \cdots j_{n}}$ in \eqref{eq:general2nF} can always be written in terms of compositions of Kronecker deltas, and for our proof it is enough to consider tensors of the form
\be
f^{i_1 \cdots i_{n}}_{j_1 \cdots j_{n}} = \delta^{i_1}_{j_1} \cdots \delta^{i_n}_{j_n}
\ee
and similar with permutations of indices. To that end, we define the flavour-nonsinglet bilinears
\be
\left( K^A \right)_i^j = \psib_i \gamma^{(A)} \psi^j \, .
\ee
Then all possible $U(\Nf)$-invariant interaction terms without derivatives take the form of linear combinations of terms of the form
\be\label{eq:JKterms}
c_{A_1 \cdots A_n} J^{A_1} \cdots J^{A_n} \quad \textrm{ or } \quad c_{A_1 \cdots A_n} \tr( K^{A_1} \cdots K^{A_n} )
\ee
or products thereof. The trace in the second term is with respect to flavour indices. The matrix-product ordering of flavour indices in this term is always possible because the components of $K^A$ are bosonic variables, hence commute with each other, and because indices on $\psi$ fields have to be contracted with indices on $\psib$ fields. By repeated application of Fierz identites, all terms of the second type in \eqref{eq:JKterms} can be reduced to terms of the first type. After a brief discussion of Fierz identities, we will prove this statement by induction on $n$.\step

{\bf Fierz identities.} \
Fierz identities relate tensor products of complex matrices with reordered indices, see {\it e.g.}~\cite{Nishi:2004st,Braun:2011pp}. In a Clifford algebra, they are a consequence of the completeness relation
\be
\frac{1}{d_\gamma} \, \gamma^{(A)}_{12}\, \gamma^{(A)}_{34} = \delta_{14}\, \delta_{32} \, ,
\ee
where Dirac indices are symbolically represented with numbers. We will need only a special case for the tensor product
\be\label{eq:gAgB}
\gamma^{(A)}_{12}\,\gamma^{(B)}_{34} = \frac{1}{d_\gamma} \, \gamma^{(C)}_{14} \left( \gamma^{(B)} \gamma^{(C)} \gamma^{(A)} \right)_{32} \, .
\ee
Because basis elements are orthogonal under the inner product
\be
\tr( \gamma^{(A)} \gamma^{(B)} ) = d_\gamma \, \delta^{AB} \, ,
\ee
any matrix $M$ in the algebra can be expressed as a linear combination of the basis elements with coefficients $c^A = \frac{1}{d_\gamma} \tr( M \gamma^{(A)} )$. This includes the product of three basis elements appearing in \eqref{eq:gAgB}. The identity \eqref{eq:gAgB} can then be expressed as
\be
\gamma^{(A)}_{12} \gamma^{(B)}_{34} = g^{BCAD} \, \gamma^{(C)}_{14} \gamma^{(D)}_{32} \, ,
\ee
where, explicitly, $g^{ABCD} = \frac{1}{d_\gamma^2} \tr( \gamma^A \gamma^B \gamma^C \gamma^D )$, although its precise form is not important for our purposes. The Fierz identities for 4F terms follow directly by contracting with fields,
\be
( K^A )_i^j \, ( K^B )_k^l = - g^{BCAD} \, ( K^C )_i^l \, ( K^D )_k^j \, .
\ee
Summing over $j = k$, we then obtain
\be\label{eq:KJreduction}
( K^A )_i^k \, ( K^B )_k^j = - g^{BCAD} \, ( K^C )_i^j \, J^D \,,
\ee
which is the key identity in proving Fierz completeness of the basis, as it allows terms of the second type in \eqref{eq:JKterms} to be reduced to the first type.\step

{\bf Proof by induction.} \
We proceed by establishing  the relation
\be\label{eq:KnProp}
c_{A_1 \cdots A_n} \tr( K^{A_1} \cdots K^{A_n} ) = \tilde{c}_{A_1 \cdots A_n} J^{A_1} \cdots J^{A_n}
\ee
for any $n \in \mathbb{N}$, and for some sets of coefficients $c_{A_1 \cdots A_n}$ and $\tilde{c}_{A_1 \cdots A_n}$ whose precise forms are not important for what follows. The proof by induction  starts with the 2F case ($n = 1$), where $\tr( K^A ) = J^A$ holds trivially. Let's also consider  the 4F case ($n = 2$), where 
\be
c_{AB} \tr( K^A K^B ) = - c_{AB} \, g^{BCAD} \, J^C J^D \, \equiv \tilde{c}_{AB} \, J^A J^B \,,
\ee
curtesy of \eqref{eq:KJreduction}. Next, we  assume that \eqref{eq:KnProp} holds true for some  $n = k>1$. Then, for $n = k + 1$, and once more using \eqref{eq:KJreduction}, we  find
\be\label{eq:n+1}
c_{A_1 \cdots A_k A_{k+1}} \tr( K^{A_1} \cdots K^{A_k} K^{A_{k+1}} )  = - c_{A_1 \cdots A_k A_{k+1}} g^{A_{k+1} B A_k C} \tr( K^{A_1} \cdots K^{A_{k-1}} K^B ) J^C \, ,
\ee
which is of the form $c'_{A_1 \cdots A_k B} \tr ( K^{A_1} \cdots K^{A_k} ) J^B$. Thankfully, the assumption that \eqref{eq:KnProp} already holds true for $n=k$ allows us to rewrite the right-hand side of \eqref{eq:n+1}  as  $c''_{A_1 \cdots A_k B} J^{A_1} \cdots J^{A_k} J^B$, which follows by considering each value of the index $B$ separately. This concludes the proof by induction, and establishes  the validity of \eqref{eq:KnProp}  for any $n \in \mathbb{N}$.\step

With this result established, it follows that all possible $U(\Nf)$ invariant  $2n$F interactions without derivatives, \eqref{eq:general2nF}, can be brought into the form 
\be\label{eq:Jfinal}
c_{A_1 \cdots A_n} J^{A_1} \cdots J^{A_n} \,.
\ee
for suitable  coefficients $c_{A_1 \cdots A_n} $. This result is used in the main text to determine the structure of  fermionic flows and their quantum effective actions at large $N$.

\section{\bf Dirac Algebra in Euclidean Signature}\label{sec:Dirac}

For convenience, we summarise our conventions for Euclidean gamma matrices and spinor bilinears. We take the Euclidean coordinates $x^\mu$ to relate to Minkowski coordinates $x_M^\mu$ in mostly minus signature as
\be
x^0 = i x_M^0 \; , \quad x^j = x_M^j \; ,
\ee
where $j = 1, \dots, d-1$. For the Euclidean gamma matrices $\gamma^\mu$, we follow  conventions as in \cite{Jakovac:2013jua,Huang:2024ypj}, 
\be\label{eq:gammaEucl}
\gamma^0 = \gamma_M^0, \quad \gamma^j = -i \gamma_M^j \; .
\ee
With this choice, the Euclidean gamma matrices are hermitian and satisfy the Clifford algebra
\be
\{ \gamma^\mu, \gamma^\nu \} = 2 \delta^{\mu \nu} \mathds{1} \; .
\ee
For the four-dimensional Clifford algebra with basis \eqref{eq:DiracBasis4d}, as well as the reducible representation \eqref{eq:DiracBasis3d} used in three dimensions, the fifth gamma matrix is defined as
\be
\gamma^5 = \gamma^0 \gamma^1 \gamma^2 \gamma^3 = i \gamma_M^0 \gamma_M^1 \gamma_M^2 \gamma_M^3 = \gamma_M^5 \, .
\ee
Altogether, analytic continuation replaces the Minkowski action $i S_M$ in the path integral by a Euclidean action $-S_E$. Specifically, if the Minkowski action takes the form
\be
S_M [ \psi, \psib ] = \int d^d x_{M} \left\{ \psib i \slashed{\partial}_M \psi  - V ( \psi, \psib ) \right\} \, ,
\ee
with interactions parametrised by a local potential $V$, the corresponding Euclidean action reads
\be
S_E [ \psi, \psib ] = \int d^d x \left\{ \psib \slashed{\partial} \psi + V ( \psi, \psib ) \right\} \, .
\ee
We refer to \cite{Wetterich:2010ni} for a more comprehensive discussion of the analytic continuation of fermionic actions between different space-time signatures.\step

Finally, we comment on our choices for spinor bilinears. Ultimately, we are interested in quantum effective actions for fermions, such as in \eqref{eq:LPAinteractions}, which at large $N$ are general functions of the bilinears. Our convention is to choose these bilinears such that they are real under the standard hermitian conjugation in Minkowski space-time, with the notion of complex conjugation extended to act on products of Grassmann numbers as \cite{DeWitt:2012mdz}
\be
\label{eq:superCC}
( z_1 \cdots z_n )^* = z_n^* \cdots z_1^* \, .
\ee 
Analogous conjugation operations can be defined in Euclidean signature, involving a reflection of the Euclidean time coordinate \cite{Wetterich:2010ni}. 
Since the effective action should be real under hermitian conjugation, the convention of having real bilinears entails that local potentials are real functions in the sense  that all of its expansion coefficients, say in a series expansion of the quantum effective action in powers of  spinor bilinears, are real.  \step

For theories with scalar and pseudo-scalar interactions, the above considerations  imply that we should work in terms of the  invariants $S = \psib\psi$ and $S_5 = i \psib \gamma^5 \psi$ in Euclidean signature, as done in \eqref{eq:SandP}, given that  their counterparts  in Minkowski space-time  $(\psib\psi)_M$ and $ (i \psib \gamma^5 \psi)_M$  are real. Similarly, for theories with vector and axial-vector interactions, the bilinears $(\psib \gamma^\mu \psi)_M$ and $(\psib \gamma^\mu \gamma^5 \psi)_M$ are real in Minkowski space-time. Continuing to Euclidean signature, we then find that these bilinears have the Euclidean counterparts $J^\mu = \psib \gamma^\mu \psi$ and $J_5^\mu =\psib \gamma^\mu \gamma^5 \psi$ via
\be
\label{eq:JsqEM}
\delta_{\mu \nu} J^\mu J^\nu = \eta_{\mu \nu} (\psib \gamma_M^\mu \psi) (\psib \gamma_M^\nu \psi) \, ,
\ee
with $\eta_{\mu\nu}$ being the  Minkowski metric, and accordingly for $J_5^2$ and $J\cdot J_5$. For these theories, the local potential only depends on the invariants $J^2, J_5^2$ and $J\cdot J_5$, and we can take it to be a real function of the Euclidean variables $J$ and $J_5$, as we have done in \eqref{eq:VandA}. Notice however that due to \eqref{eq:gammaEucl}, the relation \eqref{eq:JsqEM} carries a relative sign  that is different from the  corresponding relation for coordinate or momentum vectors, $\delta_{\mu \nu} x^\mu x^\nu = -\eta_{\mu \nu} x_M^\mu x_M^\nu$. \step

RG flows for local potential interactions are obtained by projecting the quantum effective action onto constant fields. As such, potentials depending on $n$ fermion bilinears  can be viewed as  maps from $\mathbb{R}_c^n$ to $\mathbb{R}_c$, where $\mathbb{R}_c$ denotes the algebra of real commuting (Grassmann-even) supernumbers with $z=z^*$ under \eqref{eq:superCC}. In a large-$N$ limit, where $N$ relates to the number of fermion fields,  the Grassmann algebra is infinite-dimensional. $\mathbb{R}_c$ is a subalgebra of the entire complex Grassmann algebra \cite{DeWitt:2012mdz}, and constitutes the natural setting in which differential equations for fermionic potentials should be considered. This includes the general LPA flows  \eqref{eq:LPAmain} with \eqref{eq:GinvLPA}, \eqref{eq:dK}, as well as the sample  flows \eqref{eq:SPflow} and \eqref{eq:NJLflow} for scalar- and vector-type fermionic interactions. \step

From a practical viewpoint, and in order to understand quantum effective actions as global functions of field variables, it is  useful to consider the restriction from $\mathbb{R}_c$ to $\mathbb{R}$. The Grassmann algebra contains the real numbers $\mathbb{R}$ as a subset, and standard operations of analysis work in $\mathbb{R}_c$ as  they do in $\mathbb{R}$ \cite{DeWitt:2012mdz}. This choice is also natural from the point of view of bosonisation equivalences, which relate real local potentials for scalar or pseudo-scalar bilinears in the fermionic theory to real scalar or pseudo-scalar fields in the dual Yukawa theory \cite{Weinberg:1997rv,Cresswell-Hogg:2023hdg}.

\bibliography{largeN-fermions}
\bibliographystyle{mystyle2}

\end{document}